\documentclass[journal,comsoc,final,a4paper]{IEEEtran}
\usepackage{graphicx}
\usepackage[T1]{fontenc}
\usepackage{amsmath,amsfonts,amssymb}
\usepackage[cmintegrals]{newtxmath}
\usepackage{bm}
\usepackage{algorithmic}
\usepackage{array}
\usepackage{comment}
\usepackage{stfloats}
\usepackage{bbding,soul}
\usepackage[usenames,dvipsnames]{xcolor}
\usepackage{gensymb}
\usepackage[caption=false,font=footnotesize]{subfig}
\usepackage[export]{adjustbox}
\usepackage{multirow}
\usepackage[linesnumbered,ruled,commentsnumbered]{algorithm2e}
\usepackage[hyphens]{url}

\newcommand{\overbar}[1]{\mkern 1.5mu\overline{\mkern-1.5mu#1\mkern-0.5mu}\mkern 1.5mu}

\newcommand{\specialcell}[2][c]{\begin{tabular}[#1]{@{}c@{}}#2\end{tabular}}
\newcommand{\ignore}[2]{\hspace{0in}#2}
\newtheorem{definition}{Definition}
\newcolumntype{P}[1]{>{\centering\arraybackslash}p{#1}}

\begin{document}
\title{Millimeter Wave V2V Communications: \\Distributed Association and Beam Alignment}

\author{Cristina Perfecto,~\IEEEmembership{Student Member,~IEEE}, Javier Del Ser,~\IEEEmembership{Senior Member,~IEEE,}
        \\and Mehdi Bennis,~\IEEEmembership{Senior Member,~IEEE}
\thanks{Manuscript received November 11, 2016; revised January 30, 2017; revised April 10, 2017; Accepted May 24, 2017. This work was supported by Basque Government through the ELKARTEK program (ref. KK-2016/0000096 BID3ABI), by the Academy of Finland project CARMA and by the Thule Institute strategic project SAFARI.}
\thanks{C. Perfecto and J. Del Ser are with the Department of Communications Engineering, UPV/EHU, Spain (e-mail: cristina.perfecto@ehu.eus). J. Del Ser is also with TECNALIA and with the Basque Center for Applied Mathematics (BCAM), Spain (e-mail: javier.delser@tecnalia.com).}
\thanks{M. Bennis is with the Centre for Wireless Communications, University of Oulu, Finland (e-mail: bennis@ee.oulu.fi) and also with the Department of Computer Engineering, Kyung Hee University, South Korea.}
}

\markboth{IEEE Journal on Selected Areas in Communications}%
{Perfecto \MakeLowercase{\textit{et al.}}: Millimeter Wave V2V Communication: Distributed Association and Beam Alignment}

\maketitle
\pagestyle{empty}
\thispagestyle{empty}

\begin{abstract}
Recently millimeter-wave bands have been postulated as a means to accommodate the foreseen extreme bandwidth demands in vehicular communications, which result from the dissemination of sensory data to nearby vehicles for enhanced environmental awareness and improved safety level. However, the literature is particularly scarce in regards to principled resource allocation schemes that deal with the challenging radio conditions posed by the high mobility of vehicular scenarios. In this work we propose a novel framework that blends together Matching Theory and Swarm Intelligence to dynamically and efficiently pair vehicles and optimize both transmission and reception beamwidths. This is done by jointly considering Channel State Information (CSI) and Queue State Information (QSI) when establishing vehicle-to-vehicle (V2V) links. To validate the proposed framework, simulation results are presented and discussed where the throughput performance as well as the latency/reliability trade-offs of the proposed approach are assessed and compared to several baseline approaches recently proposed in the literature. The results obtained in our study show performance gains in terms of reliability and delay up to 25\% for ultra-dense vehicular scenarios and on average 50\% more paired vehicles that some of the baselines. These results shed light on the operational limits and practical feasibility of mmWave bands, as a viable radio access solution for future high-rate V2V communications.
\end{abstract}


\IEEEpeerreviewmaketitle
\section{Introduction}
\IEEEPARstart{T}{he} last few years have witnessed the advent of wireless communications deployed in the millimeter-wave (mmWave) band, as a means to circumvent the spectrum shortage needed to satisfy the stringent requirements of 5G networks \cite{rappaport2013millimeter}. The large amount of free spectrum available in the 60 GHz band --with 14 GHz of unlicensed spectrum, roughly 15 times as much as all unlicensed Wi-Fi spectrum in lower bands-- represents a new opportunity for future communications using channel bandwidths beyond 1 GHz, as evinced by several standards for wireless personal and local area networks (such as IEEE 802.15.3c \cite{baykas2011ieee} and IEEE 802.11ad \cite{perahia2011gigabit}). This stimulating substrate for high-rate communications is the reason why 5G standardization committees and working groups are actively investing enormous research efforts towards leveraging the inherent advantages of mmWave communications (i.e. improved interference handling by virtue of highly-directive antennas) in cellular scenarios with massive device connectivity.

Among all the above scenarios where mmWave bands have been addressed in the literature, vehicular communications have lately grasped considerable attention due to more wireless technologies being integrated into vehicles for applications related to safety and leisure (infotainment), among others \cite{HeathSurvey2016}. Although certain safety applications may not require high data rates to be captured by the sensors installed in the vehicle (e.g. blind spot warning), many other applications are foreseen to require vehicular connectivity with very high transmission rates predicted to surpass the 100 Mbps limit of for raw sensor data. For instance, radars designed to operate on the 77-81 GHz band have been shown to enhance certain functionalities of vehicles such as automatic cruise control, cross traffic alert and lane change warning \cite{hasch2012millimeter}, with operating data rates far beyond the 27 Mbps limit admitted by DSRC (the \emph{de facto} standard for short-range vehicular communications \cite{Kenney:11}) or current 4G cellular communications. More advanced radar technologies such as those relying on laser technology (LIDAR) produce high-resolution maps that require even more demanding data rates (in the order of tens of Mbps, depending on the spatial resolution and scanning rate). Predictions for autonomous vehicles foresee up to 1 TB of generated data per driving hour, with rates achieving more than 750 Mbps \cite{Angelica:13}, motivating further the adoption of mmWave vehicle-to-everything (V2X) communications in the automotive sector.

Unfortunately, the challenging radio conditions derived from the mobility of vehicles, their relatively high speed with respect to pedestrians, the dynamic topology of vehicular wireless networks and its higher likelihood to produce inter-vehicular line-of-sight blockage are factors that pose significant challenges to be dealt with \cite{rappaport2014millimeter}. It has not been until recently when early findings on the propagation characteristics of mmWave vehicular communications \cite{fujise2001propagation} and limited work thereafter \cite{tsugawa2005issues} highlighted this spectrum band as a promising enabler for high-bandwidth automotive sensing \cite{Choi2016,kumari2015investigating} or beamforming in vehicle-to-infrastructure (V2I) communications \cite{va2016beam}. Interestingly, to the best of our knowledge the literature on mmWave vehicle-to-vehicle (V2V) communications is so far limited to \cite{HeathSurvey2016}, where the impact of directionality and blockage on the signal to interference plus noise ratio (SINR) are explored via simulations for unicast V2V transmissions over the 60 GHz band. However their solution is based on a static vehicle association and they do not study the delay and reliability performance associated to data traffic arrivals in the system.
\begin{table*}[t!]
	\centering
	\begin{minipage}{.95\textwidth}
		{\scriptsize
			\renewcommand{\arraystretch}{1.2}
			\caption{Summary of Notations}\label{tab:Notation}
			\centering
			\begin{tabular}{|p{1.5cm}|p{6.3cm}|p{1.5cm}|p{6.3cm}|}
				\hline \textbf{Symbol}&\textbf{Description}&\textbf{Symbol}&\textbf{Description}\\
				\hline
				$\mathcal{T}_t$, $T_t$ &Set of transmission slots, transmission slot duration.
				&$P_s$ & Packet size.\\
				$\mathcal{T}_s$, $T_s$ &Set of scheduling slots, scheduling slot duration.
				&$Q_i$, $\overbar Q_i$ &Queue length and Average Queue length in vTx $i$.\\
				$N$&Transmission slots comprised in a scheduling slot.
				&$Q_{max}$& Maximum buffer size.\\
				$T_p$& Pilot transmission duration.
				&$\lambda$& Mean packet arrival rate.\\
				$\mathcal{I}$& Set of vTx.
				&$\rho$&Traffic influx rate.\\
				$\mathcal{J}$& Set of vRx.
				&$\mathbf{A}_{\mathcal{I}}$& Random packet arrival vector in packets.\\
				$\mathcal{L}$& Set of mmWave V2V links.
				&$\mathbf{H}_{\mathcal{J}}$& Aggregate global CSI vector.\\ 
				$\ell_{i,j}$& Link between vTx $i$ and vRx $j$.
				&$\mathbf{Q}_{\mathcal{I}}$&Aggregate global QSI vector.\\
				$g_{i,j}^c$ & Channel gain in link $\ell_{i,j}$.
				&$\mathcal{X}$&Global system state.\\				
				$\delta_{i,j},\beta_{i,j}$& Channel model parameters.
				&$\Upsilon$&Global system state space.\\					
				$s_{i,j}$& Length of $\ell_{i,j}$.
				&$D_{\max}^\lambda$&Maximum latency constraint.\\
				$g_{i,j}^{t_x}$, $g_{i,j}^{r_x}$ & Antenna gain in vTx and vRx ends of $\ell_{i,j}$.
				&$P_i^p, d_{i,j}^p$&$p$-th packet in vTx $i$ queue and associated delay.\\				
				$G, g_\sphericalangle$ & Antenna mainlobe and sidelobe gains.
				&$t_{i}^{p,arr}$&Arrival time of packet $P_i^p$ in vTx $i$ queue.\\				
				$\theta_{i,j}^{t_x}$, $\theta_{i,j}^{r_x}$ & Alignment error.
				&$t_{i,j}^{p,serv}$&Departure time of $P_i^p$ (last bit) from vTx $i$ queue.\\
				$\varphi_i^{t_x}$, $\varphi_i^{r_x}$ & Beam-level beamwidths of vTx and vRx in $\ell_{i,j}$.
				&$\overbar{D}_{i,j}$&Average delay per packet per transmission slot in vTx $i$ queue.\\
				$\psi_i^{t_x}$, $\psi_i^{r_x}$ & Sector-level beamwidths of vTx and vRx in $\ell_{i,j}$.
				&$\overbar{D}_{i,j}^{sch}$&Average delay per packet per scheduling slot in vTx $i$ queue.\\
				$\tau_{i,j}$& Beamtraining associated alignment delay.
				&$\mathcal{A}_{i}^\checkmark$\hspace{-.5mm}, $\mathcal{A}_{i}^\times$&Set of successfully delivered packets from vTx $i$ queue (corr. dropped packets).\\
				$Z$ & Number of simultaneously transmitting V2V pairs.
				&$\Gamma_i^\checkmark$\hspace{-.5mm}, $\Gamma^\times_{i}$&Successfully delivered and dropped packet ratios in vTx $i$ queue.\\
				$p_{i}$ & Transmission power of reference vTx.
				&$\mathbf{\Phi}$&Matrix of possible vTx$\mapsto$vRx mappings during slot $t_s$.\\				
				$p_{z}$ & Transmission power of interfering vTx.
				&$\phi_{i,j}$&Binary association variable.\\
				$N_0$ & Gaussian background noise power density.
				&$U_{vTx}^{i,j}$,$U_{vRx}^{j,i}$&Utility of vRx $j$ when matched to vTx $i$ and viceversa.\\
				$B$ & Channel bandwidth in mmWave band.
				&$\omega_{vTx}^{i,j}$\hspace{-.5mm}, $\omega_{vRx}^{j,i}$&$\alpha$-fairness weights for vTx $i$ and vRx $j$.\\
				\hline
			\end{tabular}
		}
	\end{minipage}\hfill
\end{table*}

This work can be framed within mmWave V2V communications under the scope of Ultra-Reliable, Low-Latency Communications (URLLC), which refer to transmission technologies allowing for stringently bounded end-to-end latencies within the order of milliseconds and packet error rates on the order of $10^{-5}$ to $10^{-9}$ \cite{durisi2015towards}. Such operational limits could correspond to critical safety information captured by vehicle sensors, likely to be shared among nearby cars for an enhanced reactivity of cars against unexpected eventualities in the road. In this context we face the challenge of guaranteeing stringent latency and reliability levels in a V2V communication scenario considering the dynamic topology entailed by the movement of vehicles. Our goal is to address this challenging problem through a cross-layer information aware (CSI+QSI) vehicle association and mmWave beamwidth optimization scheme, where CSI (\emph{Channel State Information}) indicates the transmission opportunity and QSI (\emph{Queue State Information}) reflects the traffic urgency. The proposed Radio Resource Management (RRM) scheme is comprehensive and considers aspects such as the directionality (steering) of the mmWave link, the effect of the selected beamwidths on the interference at the vehicular receivers, the blockage of intermediate vehicles, the throughput versus alignment delay trade-off, the vehicle density and the impact of the speed offset between vehicles on the beam coherence time.

From the algorithmic point of view we first define utility functions that capture all the above aspects, which lay the basis for a matching game \cite{Roth1992} to solve the association problem between transmitting and receiving vehicles in a distributed fashion. Beamwidth optimization, on the other hand, is addressed using Swarm Intelligence, a class of nature-inspired optimization algorithms that simulate the collective behavior observed in certain species so as to discover optimum regions within complex search spaces under a measure of global fitness \cite{blum2008swarm}. The performance of our proposed RRM scheme is analyzed and discussed over a comprehensive set of experiments, aimed not only at exploring the quantitative performance obtained under different setups and parameters of the underlying vehicular scenario, but also as a comparison with several baselines, such as minimum-distance matching and novel pairing schemes reported in \cite{HeathSurvey2016}.

The rest of this manuscript is structured as follows: in Section \ref{sec:system_model} we describe the overall system model of the vehicular setup under consideration, and formulate the optimization problem. Section \ref{sec:scheme} and subsections therein delve into the proposed resource allocation procedure, including the adopted techniques for vehicle pairing and beamwidth optimization. In Section \ref{sec:results} we evaluate the performance of different configurations of the proposed solution under diverse settings of the considered vehicular scenario. Finally, Section \ref{sec:conclusions} concludes the paper by identifying future research directions.

\textit{Notations}: The main symbols used throughout the paper are summarized in Table \ref{tab:Notation}. Therein onwards the following notation applies: lowercase/uppercase symbols represent scalars, boldface symbols represent vectors and calligraphic uppercase symbols denote sets. The cardinality of a set is denoted by $|\cdot|$.
\section{System Model and Problem Statement} \label{sec:system_model}
This section elaborates on the system model for mmWave V2V communications, introduces the main elements that govern the cross-layer RRM policy and formulates the optimization problem that models the allocation of resources, namely, V2V links and their corresponding transmitting and receiving beamwidths.

\begin{figure}[t!]
	\centering						
	\includegraphics[width=0.8\columnwidth]{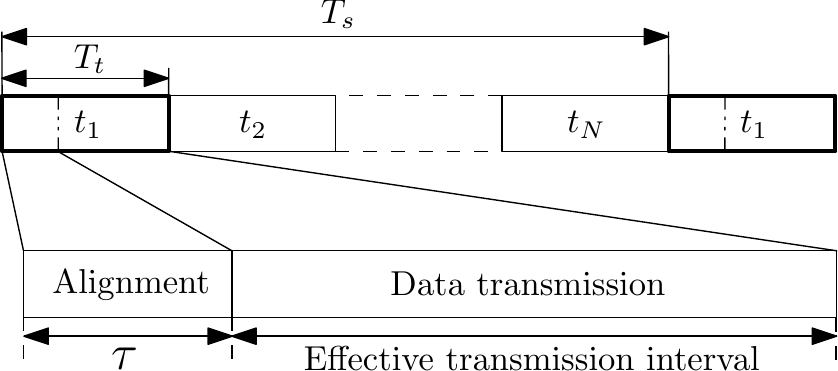}
	\caption{Detailed view of the first transmission slot within a scheduling period, divided into alignment and effective data transmission.}
	\label{fig:alignmentDelay}						
\end{figure}
\subsection{Network Topology}\label{subsec:topolAndTimeScale}

We consider a multiple lane highway road section where vehicles move at variable speeds in the same direction. Vehicles in the highway incorporate vehicular user equipments (vUEs), further separated into vehicular transmitters (vTx) and vehicular receivers (vRx), which communicate through V2V links established on mmWave frequency band operating under Time Division Duplexing (TDD). A co-channel deployment with bandwidth $B$, uniform transmit power and half-duplex mode are assumed. Let $\mathcal{I}\triangleq\{1,\ldots,I\}$, $\mathcal{J}\triangleq\{1,\ldots,J\}$ and $\mathcal{L}\triangleq\{1,\ldots,L\}$, with $\mathcal{I}\cap\mathcal{J}=\emptyset$, $\mathcal{|L|}\leq \min\{\mathcal{|I|},\mathcal{|J|}\}$ respectively denote the sets of vTx, vRx and links in the system.

In this scenario the relative movement between vehicles causes a varying network topology with changing channel conditions, misalignments between vehicle pairs and uncontrollable blocking effects in the deployed millimeter-wave links. This strong topological variability and the increased complexity of instantaneous, uncoordinated RRM policies impose the need for time-slotted communications, with two different time scales:
\begin{itemize}
	\item Data transmission slots (ms) denoting the intervals $[t,t+T_t)$, with $T_t$ as the duration of the transmission period.
	\item Scheduling slots (ms) which hereafter refers to the intervals $[t,t+T_s)$, with $T_s$ representing the duration of the network-wide enforced control actions.	
\end{itemize}

Without loss of generality, each scheduling slot is assumed to comprise an integer number $N$ of transmission slots (i.e. $T_s=NT_t$) such that scheduling occurs at $\mathcal{T}_s\triangleq\{t_s\in \mathbb{N}: t_s\mod{N}=0\}$, and data transmission is held at $\mathcal{T}_t \triangleq \mathbb{N}$. As shown in Fig. \ref{fig:alignmentDelay} the initial transmission slot within a scheduling slot in $\mathcal{T}_s$ will be further divided into two phases: 1) the antenna steering or beam alignment phase, whose duration depends on the beamwidths selected at each vTx/vRx pair; and 2) the effective data transmission phase, which starts once boresight directions have been correctly aligned. This split will only hold at those time intervals where a new scheduling policy is triggered and deployed.
\subsection{Channel Modelling}

To model the 60 GHz mmWave channel and simultaneously account for blockage effects on the mmWave signal, the standard log-distance pathloss model proposed in \cite{Yamamoto2008} is adopted. Under this model the channel gain $g_{i,j}^{c}$ on link $\ell_{i,j}$ between vTx $i$ and vRx $j$ is given by
\begin{equation}\label{eq:pathloss}
g_{i,j}^{c}= 10\:\delta_{i,j}\log_{10}(s_{i,j})+\beta_{i,j}+15\:s_{i,j}/1000,
\end{equation}
where the third term represents the atmospheric attenuation at 60 GHz, and the values for parameters $\delta_{i,j}$ --the pathloss exponent-- and $\beta_{i,j}$ depend on the number of blockers that obtrude the link connecting a given vTx $i$ with its corresponding pair vRx $j$. The original model in \cite{Yamamoto2008} was recently generalized in \cite{HeathSurvey2016} by providing values for $\delta$ and $\beta$ when the number of blocking vehicles goes beyond three. Since we deal with a dynamic scenario, the channel gain will vary along time as a result of the relative movement of the vehicles, which yields $g_{i,j}^{c}(t)$. At the end of any given transmission slot $t\in\mathcal{T}_t$, the aggregate global CSI\footnote{Instantaneous reporting of CSI and QSI related side effects (e.g. increased signaling overhead) will be avoided by enforcing a long-term RRM strategy that includes, among others, learning techniques.} for the set of $|\mathcal{J}|$ receivers will be given by $\mathbf{H}_{\mathcal{J}}(t)=\{H_{j}(t):\forall j\in\mathcal{J}\}$, with $H_{j}(t)=g_{i,j}^{c}(t)$ if link $\ell_{i,j}$ exists.

\begin{figure}[t!]
	\begin{tabular}{@{}l@{}}
		\subfloat[]{
			\includegraphics[width=.4\columnwidth,height=0.985\columnwidth]{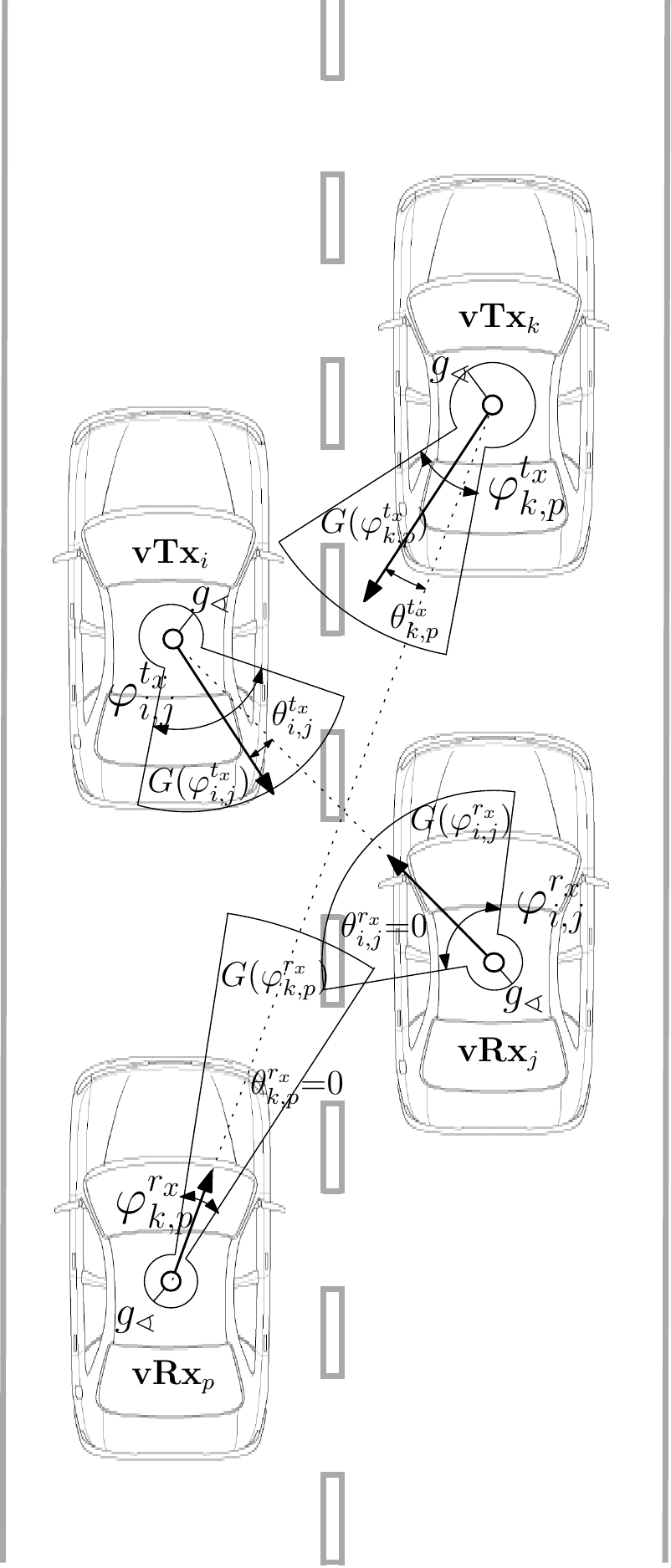}
			\label{fig:alignGeneral}}
	\end{tabular}
	\begin{tabular}{{@{}c@{}}}
		\vspace{-1.8mm}
		\subfloat[]{
			\includegraphics[scale=.43]{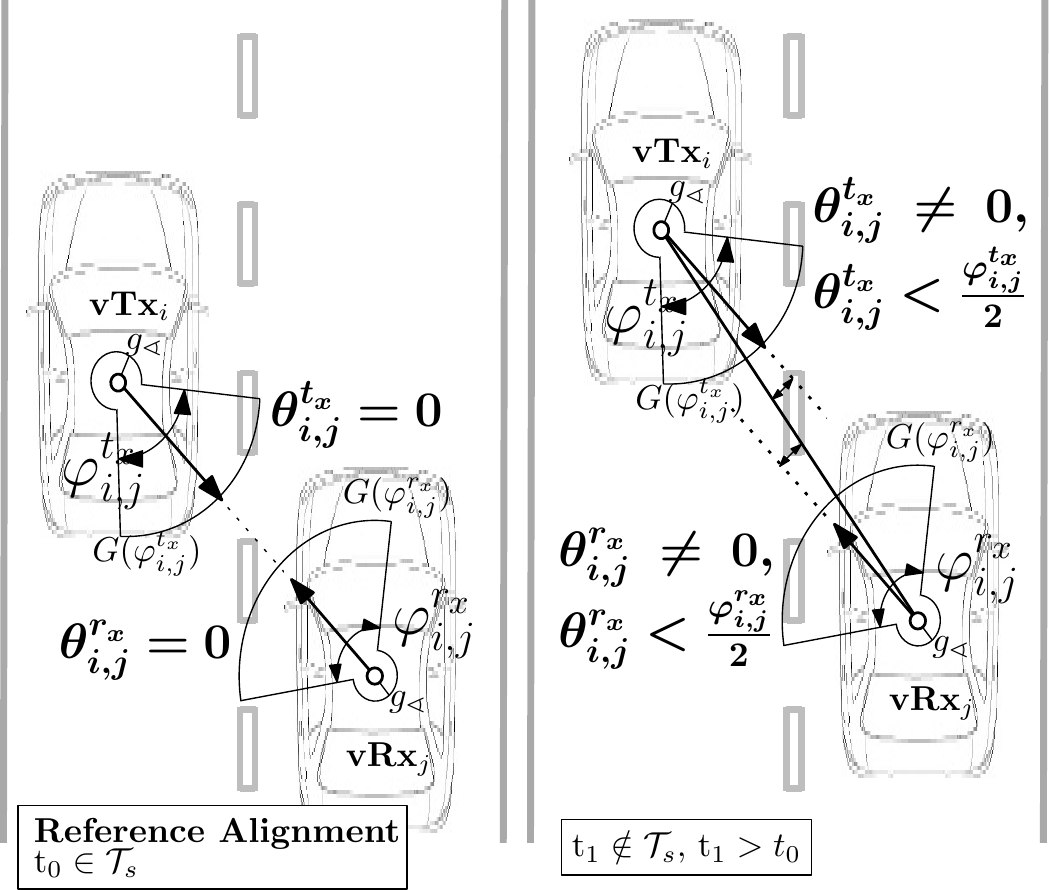}
			\label{fig:misalignWide}}\\
		\subfloat[]{
			\includegraphics[scale=.43]{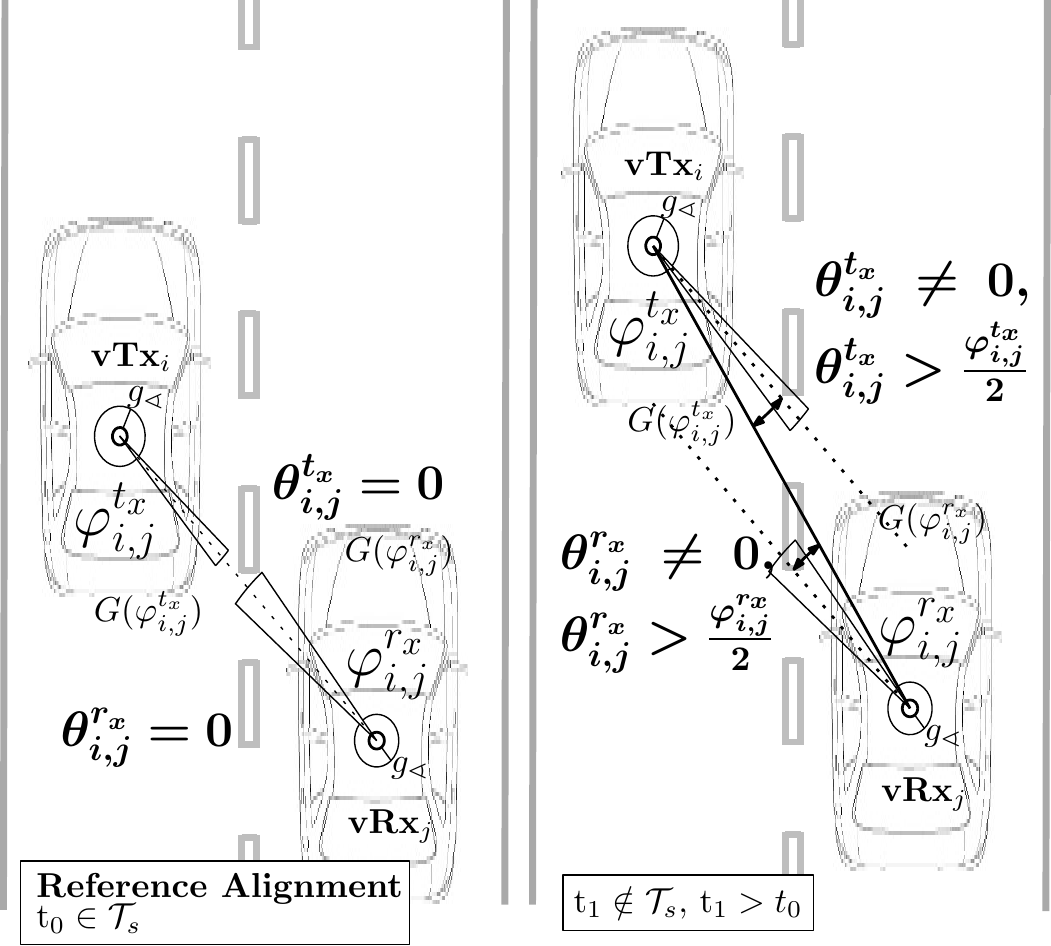}
			\label{fig:misalignNarrow}}
	\end{tabular}
	\caption{(a) Parameters of the ideal sectored antenna model under study. Effect of the misalignment between transmitter and receiver boresight directions on the vTx and vRx antenna gains with (b) wide and (c) narrow beamwidths.}
	\label{fig:angles}		
\end{figure}
\subsection{Antenna Pattern}

For the sake of tractability directional antenna patterns in vehicles will be approximated by a two-dimensional ideal sectored antenna model as represented by Fig. \ref{fig:angles}(a). This model captures the four most relevant features of the radiation pattern, namely the boresight direction, the directivity gains in the mainlobe and in the sidelobe (also referred to as front-to-back ratio) and the half-power beamwidth. Transmission and reception directivity gains $g_{i,j}^\wp(t)$ ($\wp\in\{t_x,r_x\}$) of vehicles in link $\ell_{i,j}$ during a transmission slot $t\in\mathcal{T}_t$ are given by \cite{Wildman:14}
\begin{equation}\label{eq:antenna}
g_{i,j}^\wp(t)\hspace{-1mm}=\hspace{-1mm}\left\lbrace \hspace{-0.5mm}
\begin{array}{ll}
\hspace{-1.2mm}G\left(\varphi_{i,j}^\wp\right)\hspace{-1mm}=\hspace{-1mm}\frac{2\pi-\left(2\pi-\varphi_{i,j}^\wp(t)\right)g_\sphericalangle}{\varphi_{i,j}^\wp(t)}\text{,}\hspace{-1mm}&\hspace{-1mm}\text{if }\hspace{-0.5mm} |\theta_{i,j}^\wp(t)|\hspace{-1mm}\leq \varphi_{i,j}^\wp/2,\\
\vspace*{-0.1cm}
\hspace{-1.2mm}g_\sphericalangle,\hspace{-1mm}&\hspace{-1mm}\text{otherwise,}
\end{array}	\right.
\end{equation}
where $\theta_{i,j}^\wp(t)$ represents the alignment error between vTx$_i$ and vRx$_j$ antenna steering directions and the corresponding boresight directions of vRx$_j$ and vTx$_i$, $\varphi_{i,j}^\wp(t)$ is the half-power beamwidth of link $\ell_{i,j}$ at transmission ($\wp=t_x$) and reception ($\wp=r_x$) sides set for the scheduling period at hand, and $0\leq g_\sphericalangle\ll 1$ is the non-negligible sidelobe power. 

As exemplified in Fig. \ref{fig:angles}(b) and Fig. \ref{fig:angles}(c), the likeliness of misalignment impacting on desired links due to a non-continuous steering/beamtracking mechanism may vary depending on several factors, such as the relative speed of the vehicles involved in the link, the width of the mainlobes of the transmitter and receiver antennas, or the length of the scheduling interval. Moreover, the selected beamwidths will impel whether signals from undesired V2V links arrive into the sidelobes or the mainlobe of vRxs, which will severely impact measured SINR levels. For this reason the sought RRM should also include a beamwidth selection strategy that dynamically adapts to the surrounding conditions and, counteracts their negative effect on the transmitted signal --which, in turn, comes along with an impact on the dynamics of the transmission queues--. The latter gains relevance in realistic scenarios, where the dynamics of the vehicle movement involve frequent misalignment events.  
\subsection{Alignment Delay and Transmission Rate}

Although numerous alternatives that speed up the beamforming protocol have been proposed in the literature, such as \cite{Tsang2011} or more recently \cite{Nitsche2015,Palacios2016}, a simplified version of the three-step beam codebook-based approach introduced by \cite{Wang2009} is employed due to its robustness and compliance with ongoing standards. Specifically, a two-staged beam alignment process will yield the best steering for the refined beams at both ends of the V2V link. These two stages encompass a sequence of pilot transmissions and use a trial-and-error approach where first a coarse sector-level scan detects best sectors for vTx and vRx and, afterwards, within the limits of the selected sector a finer granularity beam-level sweep searches for best beam-level pairs. In this approach the well-known alignment delay versus throughput trade-off \cite{Shokri-Ghadikolaei2015} is exposed: the selection of narrower beamwidths induces longer training overheads and yields reduced effective transmission rates.

Without loss of generality we assume here that for each vehicle in a V2V link before the beam-level alignment phase itself, either the sector level alignment has already been performed or that coarse location of neighboring vehicles has been learned (e.g. during the learning process in Section \ref{subseq:learning}), effectively reducing the beam search. By applying a continuous approximation \cite{Shokri-Ghadikolaei2015}, the alignment time penalty $\tau_{i,j}(t)$ can be quantified as 
\begin{equation} \label{eq:tau}
\tau_{i,j}(t)\triangleq\tau_{i,j}\left(\varphi_{i,j}^{t_x}(t),\varphi_{i,j}^{r_x}(t)\right)=\frac{\psi_i^{t_x}\psi_j^{r_x}}{\varphi_{i,j}^{t_x}(t)\varphi_{i,j}^{r_x}(t)}{T_p},
\end{equation}
where $\psi_i^{t_x}$ and $\psi_j^{r_x}$ denote the sector-level beamwidths of vTx $i$ and vRx $j$, and $T_p$ denotes the pilot transmission duration. Constraints coming from the operational array antenna limits, sector level beamwidths and the fact that $\tau_{i,j}(t)$ should not exceed $T_t$ restrict the values taken by the vTx and vRx beamwidths $\varphi_{i,j}^{t_x}(t)$ and $\varphi_{i,j}^{r_x}(t)$, i.e.
\begin{equation}\label{eq:low_beamwidth_product}
\varphi_{i,j}^{t_x}(t)\varphi_{i,j}^{r_x}(t)\geq\frac{T_p}{T_t}{\psi_{i}^{t_x}}{\psi_{j}^{r_x}}.
\end{equation}

Under these assumptions the maximum achievable data rate $r_{i,j}(t)$ between vTx $i$ and vRx $j$ will depend on whether beam alignment is performed at time slot $t$ with its corresponding induced delay and on the measured SINR at vRx $j$, including the interference of other incumbent vTxs on vRx $j$. The rate for a time slot $t$ of duration $T_t$ over which alignment is performed, is given by
\begin{equation} \label{eq:rate}
r_{i,j}(t)=\left(1-\frac{\tau_{i,j}(t)}{T_t}\right)B\log_2\left(1+\text{SINR}_{j}(t)\right),
\end{equation}
where the SINR at time slot $t$ under $Z=|\mathcal{Z}|$ simultaneously transmitting vTxs is given by
\begin{equation} \label{eq:sinr}
\text{SINR}_{j}(t)=\frac{p_ig_{i,j}^{t_x}(t)g_{i,j}^c(t)g_{i,j}^{r_x}(t)}{\sum\limits_{\substack{z\in\mathcal{Z}\subseteq\mathcal{I}\\z\neq i}} p_zg_{z,j}^{t_x}(t)g_{z,j}^c(t)g_{z,j}^{r_x}(t) +N_0B},
\end{equation}
with $p_{i}$ being the transmission power of reference vTx $i$; $g_{i,j}^c(t)$ the channel gain in the link $\ell_{i,j}$; $g_{i,j}^{t_x}(t)$ and $g_{i,j}^{r_x}(t)$ respectively denoting the antenna gains at the transmitting and receiving ends of the link. The leftmost term $p_zg_{z,j}^{t_x}(t)g_{z,j}^c(t)g_{z,j}^{r_x}(t)$ in \eqref{eq:sinr} represents the contribution of the interference received at vRx $j$ from vTx $z$, $\forall z\in\mathcal{Z}\subseteq\mathcal{I}, z\neq i$; while in the rightmost term, $N_0$ is the Gaussian background noise power density (dBm/Hz) and $B$ is the bandwidth of the mmWave band. Finally, it is also straightforward to note that the rate $r_{i,j}(t)$ increases when no alignment is performed during the time slot $t$, as per \eqref{eq:rate} with $\tau_{i,j}(t)=0$.
\subsection{Queues and Delay Modeling}\label{subsec:queues}

Since our target is to design a adaptive RRM policy appropriate for a delay-sensitive information flow, a model that captures the traffic and queue dynamics is needed. For this purpose each vTx will maintain a queue for data that arrives from upper layers of the protocol stack.Assuming a fixed packet size $P_s$ in bits, let $Q_i(t)$ be the queue length in number of packets of vTx $i$ matched to vRx $j$ at the beginning of time slot~$t$. Let $\mathbf{A}_{\mathcal{I}}(t)=(A_1(t),...,A_I(t))$ denote the random packet arrivals vector (in number of packets) to the set $\mathcal{I}$ of vTxs at the end of time slot $t\in\mathcal{T}_t$ i.e., new arrivals are observed after the scheduler's action has been performed. We assume that every entry $A_i(t)$ in $\mathbf{A}_{\mathcal{I}}(t)$, $\forall i\in \{1,\ldots,I\}$, is independently and identically distributed (i.i.d) over time slots due to mutually independent packet arrival processes following a Poisson distribution with mean $\mathbf{E}[A_i(t)]=\lambda$ within the stability region of the system. Then, if the rate in $\ell_{i,j}$ is $r_{i,j}(t)$ as per \eqref{eq:rate}, a maximum of $r_{i,j}(t)T_t/P_s$ packets will be successfully transmitted during slot $t\in\mathcal{T}_t$, and the queue dynamics for vTx $i$ are given by
	\begin{equation} \label{eq:queuedynamics}
	Q_i(t + 1) = \min\left\lbrace\left(Q_i(t)-\frac{r_{i,j}(t)T_t}{P_s}\right)^+ +A_i(t),Q_{\max}\right\rbrace,
	\end{equation}
with $Q_i(t)\in \mathbb{R}$, $x^+ \triangleq \max\{x,0\}$, and $Q_{\max}$ the maximum buffer size of the queue. With this notation, we let $\mathbf{Q}_{\mathcal{I}}(t)=\{Q_{i}(t):\forall i\in\mathcal{I}\}$ represent the aggregate global QSI vector for the set $\mathcal{I}$ of vTxs at the beginning of time slot $t\in\mathcal{T}_t$. Finally, we define the global system state at time slot $t\in\mathcal{T}_t$ as $\mathcal{X}(t)\triangleq(\mathbf{H}_{\mathcal{J}}(t),\mathbf{Q}_{\mathcal{I}}(t))\in \Upsilon$, with 
$\Upsilon$ denoting the global system state space.

Upon its arrival to a certain queue, a packet will be either delivered or dropped within $D_{\max}^\lambda$ ms after entering the queue:
\begin{itemize}
	\item If link $\ell_{i,j}$ is active and channel conditions in the link are good enough, packet $P_i^p$ (with $p\in\{1,\ldots,A_i(t)\}$) will be transmitted with a delay $d_{i,j}^p\le D_{\max}^\lambda$ given by 
		\begin{equation}\label{eq:delay_packet}
		d_{i,j}^p = t_{i,j}^{p,serv}-t_{i}^{p,arr},
		\end{equation}
	with $t_{i}^{p,arr}$, $t_{i,j}^{p,serv}$ respectively denoting the arrival time of packet $P_i^p$ at the queue and the time when the last of the bits of $P_i^p$ is transmitted to vRx $j$ i.e, $d_{i,j}^p$ is a joint measure of queue waiting time and transmission delay\footnote{By a slight abuse in the notation, we keep subindex $j$ in $t_{i,j}^{p,serv}$ and related delay statistics to explicitly refer to the dependence of such terms on the transmission rate $r_{i,j}(t)$ of the channel from vTx $i$ to its paired vRx $j$.}. In general, the average delay per packet $\overbar{D}_{i,j}(t)$ during transmission slot $t\in\mathcal{T}_t$ can be computed by averaging the delays $d_{i,j}^p$ of each packet successfully delivered over this link for the slot at hand, as
		\begin{equation}\label{eq:delay_transmission_comp}
		\overbar{D}_{i,j}(t) = \frac{\sum_{p\in \mathcal{A}_{i}^\checkmark(t)} d_{i,j}^p}{\left|\mathcal{A}_{i}^\checkmark(t)\right|},
		\end{equation}
	where $\mathcal{A}_{i}^\checkmark(t)$ denotes the subset of packets successfully sent towards vRx $j$ at time $t\in \mathcal{T}_t$. From this definition the average delay per delivered packet over the scheduling period $t_s\in\mathcal{T}_s$ will be given by
		\begin{equation}\label{eq:delay_scheduling_comp}
		\overbar{D}_{i,j}^{sch}(t_s) = \frac{\sum_{t=t_s-N+1}^{t_s} \overbar{D}_{i,j}(t)}{N}.
		\end{equation}
	
	\item If link $\ell_{i,j}$ is active but channel conditions in the link are not good enough to deliver pending packets towards receiver vRx $j$ within $D_{\max}^\lambda$ and, either a new traffic arrival event is triggered at transmitter vTx $i$ or a new scheduling slot starts, unfinished packets will be dropped from the queue. In both cases, the rationale behind the adoption of such a hard requirement is to prioritize newer traffic and to ensure minimum-delay communications. Each time a packet is dropped, a penalty will be incurred and computed in the form of reliability loss. This modeling is often adopted in the context of URLLC \cite{3GPP:URLLC}. Specifically, the set of dropped packets in a transmission slot $t\in\mathcal{T}_t$ will be denoted as $\mathcal{A}_{i}^\times(t)$, such that both $\mathcal{A}_{i}^\times(t) \cap \mathcal{A}_{i}^\checkmark(t)=\emptyset$ and $|\mathcal{A}_{i}^\times(t)| \cup |\mathcal{A}_{i}^\checkmark(t)|\leq Q_{i}(t)$ are met. Finally, the packet dropping ratio is defined at the scheduling slot level as 
		\begin{equation} \label{eq:packet_drop_ratio_sch}
		\color{black}\Gamma^\times_{i}(t_s)\hspace{-.5mm}\triangleq\hspace{-.5mm} \frac{\sum_{t=t_s-N+\hspace{-.1mm}1}^{t_s}\hspace{-.5mm}\left|\mathcal{A}_{i}^\times(t)\right|}{\sum_{t=t_s-N\hspace{-.1mm}+1}^{t_s}A_{i}(t)}\hspace{-.5mm}=\hspace{-.5mm}1\hspace{-.5mm}-\frac{\hspace{-.8mm}\sum_{t=t_s-N+\hspace{-.1mm}1}^{t_s} \hspace{-.5mm}\left|\mathcal{A}_{i}^\checkmark\hspace{-.5mm}(t)\right|}{\sum_{t=t_s-N+\hspace{-.1mm}1}^{t_s}A_{i}(t)}.
		\end{equation}
\end{itemize}
\vspace*{-8mm}
\subsection {Elements of RRM and Problem Statement}\label{subsec:mainP}
In order to formally define an RRM policy we let $\bm{\Phi}(t_s)\triangleq\{\phi_{i,j}(t_s): i\in\mathcal{I}(t_s),\:j\in\mathcal{J}(t_s)\}$ denote the set of all possible vTx/vRx mappings in the system in a given scheduling slot $t_s\in \mathcal{T}_s$. Note here that $\mathcal{I}(t_s)$ (corr. $\mathcal{J}(t_s)$) denotes the subset of vTx and vRx present on the road scenario at scheduling time $t_s$. \ignore{By a slight abuse in notation} We further define $\mathcal{I}_j(t_s)\subseteq \mathcal{I}(t_s)$ and $\mathcal{J}_i(t_s)\subseteq \mathcal{J}(t_s)$ as the subsets of feasible vTxs for vRx $j$ and the feasible vRxs for vTx $i$, where feasibility is due to a circular coverage constraint of radius $R_c$ (in meters). In this set $\phi_{i,j}(t_s)$ will represent the association variable so that for the pair composed by vTx $i$ and vRx $j$
\begin{equation}
\phi_{i,j}(t_s)=\left\lbrace
\begin{array}{ll}
1 &\text{if link $\ell_{i,j}$ is set, $\forall t\in[t_s,t_s+N)$,}\\
0 &\text{otherwise.}\\
\end{array}
\right.
\end{equation}

Bearing this in mind, $\bm{\Phi}(t_s)$ jointly with a proper selection of the beamwidths at both vTx and vRx as defined by
\begin{equation}
	\bm{\varphi}^{t_x}\hspace{-.5mm}(t_s)\hspace{-1mm} \triangleq\hspace{-1mm} \left\{\varphi_{i,j}^{t_x}(t_s)\hspace{-1mm}:\hspace{-.3mm}i\hspace{-.6mm}\in\hspace{-.7mm}\mathcal{I}(t_s), j\hspace{-1mm}\in\hspace{-1mm} \mathcal{J}_i(t_s)\hspace{.5mm}\text{such}\hspace{-.5mm}\text{ that}\hspace{1mm}\phi_{i,j}(t_s)\hspace{-1mm}=\hspace{-1mm}1\right\}\hspace{-.3mm},
\end{equation}
\begin{equation}
	\bm{\varphi}^{r_x}\hspace{-.5mm}(t_s)\hspace{-1mm} \triangleq\hspace{-1mm} \left\{\varphi_{i,j}^{r_x}(t_s)\hspace{-1mm}:\hspace{-.5mm}j\hspace{-.6mm}\in\hspace{-.7mm}\mathcal{J}\hspace{-.5mm}(t_s), i\hspace{-1mm}\in\hspace{-1mm} \mathcal{I}_j(t_s)\hspace{.5mm}\text{such}\hspace{-.5mm}\text{ that}\hspace{1mm}\phi_{i,j}(t_s)\hspace{-1mm}=\hspace{-1mm}1\right\}\hspace{-.3mm},
\end{equation}
give rise to the effective instantaneous rate $r_{i,j}(t,\bm{\Phi}(t_s))$ of link $\ell_{i,j}$, as per Expressions \eqref{eq:rate} and \eqref{eq:sinr} with $\mathcal{Z}=\mathcal{I}(t_s)$ and relative interferences and gains between pairs given by the prevailing matching policy $\bm{\Phi}(t_s)$. Namely,
\begin{equation}
r_{i,j}(t,\bm{\Phi}(t_s))\hspace{-0.8mm}=\hspace{-0.8mm}\left(1-\frac{\tau_{i,j}(t)}{T_t}\right)B\log_2\left(1\hspace{-0.8mm}+\hspace{-0.8mm}\text{SINR}_{j}(t,\bm{\Phi}(t_s))\right) \mbox{},
\end{equation}
if $t=t_s$ (i.e. the first transmission slot after scheduling at time $t_s\in\mathcal{T}_s$ has been enforced), while for $t\in[t_s+1,t_s+N)$,
\begin{equation}\label{eq:rate_with_matching}
r_{i,j}(t,\bm{\Phi}(t_s))\hspace{-0.8mm}=B\log_2\left(1\hspace{-0.8mm}+\hspace{-0.8mm}\text{SINR}_{j}(t,\bm{\Phi}(t_s))\right).
\end{equation}

Based on this rate and the traffic influx rate defined as $\rho=\lambda P_s$, a fraction of the packets generated at vTx $i$ will be transmitted towards vRx $j$, producing delays and packet dropping statistics over a given scheduling slot. For that reason a delay-sensitive RRM policy should take into account not only the finite delay of those packets successfully transmitted towards their destinations (for which queue dynamics are set to prioritize new incoming traffic), but also the interplay between delay and dropped packets enforced by the queuing policy.

The problem tackled in this work can be hence formulated as the design of the RRM policy $\{\bm{\Phi}(t_s),\bm{\varphi}^{t_x}(t_s),\bm{\varphi}^{r_x}(t_s)\}$ for $t_s\in\mathcal{T}_s$ such that
\begin{subequations} \label{eq:mainOptprb}
	\begin{align}
	&\underset{\bm{\Phi}(t_s),\bm{\varphi}^{t_x}(t_s),\bm{\varphi}^{r_x}(t_s)}{\text{Minimize}}&&\hspace{-2.2mm}\sum\limits_{i\in\mathcal{I}(t_s)}\sum\limits_{j\in\mathcal{J}(t_s)}\overbar{D}_{i,j}^{sch}(t_s)\phi_{i,j}(t_s), \label{eq:fitness}\\
	& \quad\;\text{subject to:} && \hspace{-2.2mm}\overbar{Q}_{i}(t)<\infty, \:\:\forall t\in (t_s-N,t_s], \label{eq:qinf}\\  
	&&&\hspace{-2.2mm}\sum\limits_{j\in \mathcal{J}(t_s)}\phi_{i,j}(t_s)=1, \forall i \in \mathcal{I}(t_s), \label{eq:matchres1}\\
	&&&\hspace{-2.2mm}\sum\limits_{i\in \mathcal{I}(t_s)}\phi_{i,j}(t_s)=1,  \forall j \in \mathcal{J}(t_s), \label{eq:matchres2}\\
	&&&\hspace{-2.2mm}\phi_{i,j}(t_s)\hspace{-0.8mm} \in\hspace{-0.8mm} \{0,1\}, \forall i,j\hspace{-0.8mm}\in\hspace{-0.8mm} \mathcal{I}(t_s)\hspace{-0.8mm}\times\hspace{-0.8mm}\mathcal{J}(t_s), \label{eq:matchres3}\\ 
	&&&\hspace{-2.2mm}\varphi_{i,j}^{t_x}(t_s)\:\varphi_{i,j}^{r_x}(t_s)\hspace{-1mm}\geq\hspace{-1mm} \frac{T_p}{T_t}\psi_{i,j}^{t_x}\psi_{i,j}^{r_x},\label{eq:mainOptprb_f} \\
	&&&\hspace{-2.2mm}\varphi_{i,j}^{t_x}(t_s)\leq \psi_{i,j}^{t_x}, \label{eq:mainOptprb_g}\\
	&&&\hspace{-2.2mm}\varphi_{i,j}^{r_x}(t_s)\leq\psi_{i,j}^{r_x}, \label{eq:mainOptprb_h}
	\end{align}
\end{subequations}
where inequality \eqref{eq:qinf} indicates that no queue should overflow during the scheduling period at hand; Expressions \eqref{eq:matchres1} through \eqref{eq:matchres3} denote that vehicles are paired one-to-one; and inequalities \eqref{eq:mainOptprb_f} through \eqref{eq:mainOptprb_h} reflect the bounds imposed on the beamwidths to be allocated as per \eqref{eq:low_beamwidth_product}.

The above optimization problem is difficult to solve analytically and is computationally hard, especially in vehicular environments calling for low-complexity distributed solutions. For this reason we will decompose it into two problems: the vehicle pairing and the beamwidth optimization. Subsequently, tools from Matching Theory and from Swarm Intelligence are leveraged to account, respectively, for the optimization of $\bm{\Phi}(t_s)$, and the selection of the beamwidths of both sides of each established mmWave V2V link (corr. $\bm{\varphi}^{t_x}(t_s)$ and $\bm{\varphi}^{r_x}(t_s)$). We will then explore the operational limits in terms of $\overbar{D}_{i,j}^{sch}(t_s)$ and $\Gamma^\times_{i}(t_s)$ under different scheduling interval durations, traffic packet arrival rates and packet sizes. The ultimate goal of this study is to numerically assess the \emph{reliability} of different RRM policies in mmWave V2V communications defined as the ratio of the number of packets $\Gamma_i^\checkmark (t_s)\triangleq 1-\Gamma_i^\times (t_s)$ of size $P_s$ successfully received at every receiver within a maximum delay $D_{max}^\lambda$\cite{Pocovi2015,Soret2014}.

\begin{figure}[t!]
	\centering
	\includegraphics[width=.99\columnwidth]{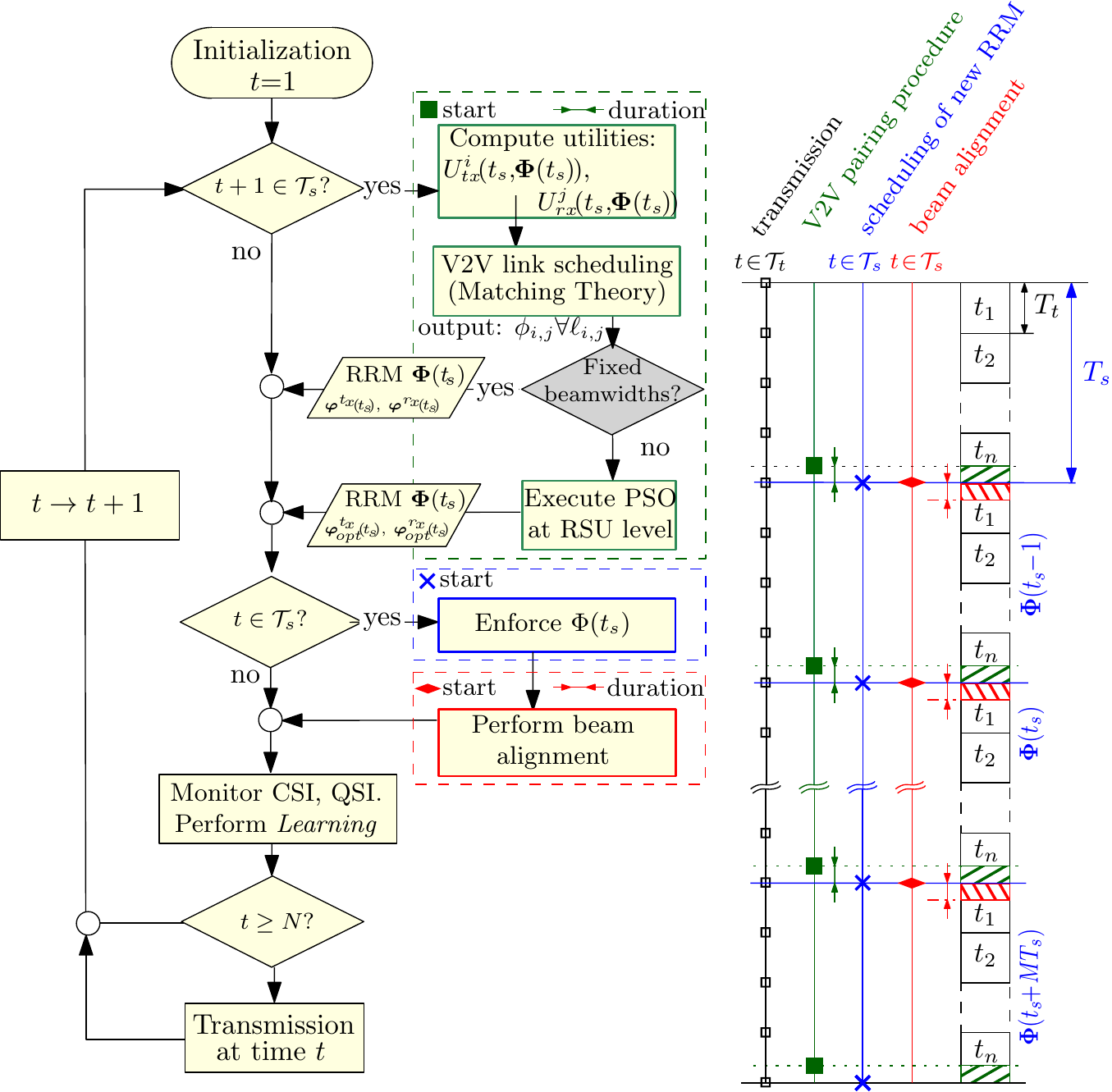}
	\caption{Interrelation between processes performed at different time scales.}
	\label{fig:timeScale-flowchart}
	\vspace{-0.5cm}							
\end{figure}
\section{Proposed Scheme} \label{sec:scheme}

Our objective in this work is to design a self-organizing mechanism to solve the vehicle-to-vehicle association problem, in a decentralized manner, in which  vTxs and vRxs interact and decide to link to each other based on their utilities. To this end, Matching Theory \cite{Roth1992}, a Nobel Prize winning framework, offers a promising approach for resource management in wireless communications \cite{Gu2015}. As depicted schematically in Fig. \ref{fig:timeScale-flowchart}, elements from Matching Theory are used for allocating mmWave V2V links in the setup at every scheduling slot $t_s$, with a previous learning process to capture essential information required for the matching game. Learning and matching are then followed by an optimization phase that allocates transmission and reception beamwidths for the matched pairs. Finally, beam alignment is performed. Prior to defining the matching game itself, we will first introduce the framework and specify the utility functions for both sets of agents, as well as the learning process upon which utilities will be computed.
\LinesNumberedHidden{
	\begin{algorithm}[t]
		\scriptsize
		\caption{Proposed CSI/QSI-aware V2V Matching Algorithm}
		\label{algo:DA-v2v}
		\DontPrintSemicolon 
		\KwData{Just before $t=t_s$, $\forall t_s\in\mathcal{T}_s$: All vRxs and vTxs are unmatched, i.e. $\forall i\in\mathcal{I}(t_s)$, $\forall j\in\mathcal{J}(t_s)$,  $\phi_i(t_s)=\emptyset$, $\phi_j(t_s)=\emptyset$).}
		\KwResult{Convergence to a stable matching $\boldsymbol{\Phi}(t_s)$.}
		\textbf{Phase I - Information exchange;}\\
		\begin{itemize}
			\item Each vRx $j$ sends to vTxs on its vicinity, i.e. $\mathcal{I}_j(t_s)$, entries $\{t',i,\mbox{SINR}_j(t')\}$ collected from pilot transmissions for link exploration.
			\item $\overbar{r}_{i,j}^{est}(t_s)$ is computed as per \eqref{eq:rateestimated}. Estimated QSI is computed as $\overbar{Q}_{i,j}(t_s)=P_s/\overbar{r}_{i,j}^{est}(t_s)$. 
		\end{itemize}
		\textbf{Phase II - Matching game construction;}
		\begin{itemize}
			\item Each vTx $i$, $\forall i \in \mathcal{I}(t_s)$, updates $U_{vTx}^{i,j}(t_s)$ over the $\mathcal{J}_i(t_s)$ vRxs as per \eqref{eq:utilityTx-final};
			\item Each vRx $j$, $\forall j \in \mathcal{J}(t_s)$, updates $U_{vRx}^{j,i}(t_s)$ over the $\mathcal{I}_j(t_s)$ vTxs as per \eqref{eq:utilityRx-final};
		\end{itemize}
		\textbf{Phase III - Deferred Acceptance for V2V link allocation;}
		\begin{itemize}
			\item For each vRx $j$, initialize the subset of its eligible vTxs, $\mathcal{E}_j\subseteq\mathcal{I}_j$ so that  $|\mathcal{E}_j|=|\mathcal{I}_j|$ 
			\item Initialize the subsets of unmatched vRxs  $\mathcal{S}^{Rx}\subseteq\mathcal{J}(t_s)$, and unmatched vTxs  $\mathcal{S}^{Tx}\subseteq\mathcal{I}(t_s)$ so that $|\mathcal{S}^{Rx}|=|\mathcal{J}(t_s)|$ and $|\mathcal{S}^{Tx}|=|\mathcal{I}(t_s)|$
		\end{itemize}
		with $|\cdot|$ denoting cardinality.\\
		\While{$|\mathcal{S}|\not=\emptyset \mbox{ and} \sum_{j \in \mathcal{S}^{rx}}|\mathcal{E}_j|\not=\emptyset $}{
			Pick a random vRx $j\in\mathcal{S}^{rx}$;\\
			\If{$|\mathcal{E}_j|\not=\emptyset$}{
				vRx $j$ sends V2V link proposal to its best ranked vTx $n$, $n\in\mathcal{E}_j$;\\
				\If{$n\in\mathcal{S}^{tx}$}{ 
					Match $j$ and $n$ setting $\phi_j(t_s)=n$ and $\phi_n(t_s)=j$;\\
					Remove $j$ and $n$ from $\mathcal{S}^{rx}$ and $\mathcal{S}^{tx}$ respectively;\\
				}
				\Else{ 
					\If{$U_{vTx}^{n,j}>U_{vTx}^{n,\phi_n(t_s)}$}{ 
						Reject proposal from $\phi_n(t_s)$; add back $\phi_n(t_s)$ to $\mathcal{S}^{rx}$ and remove $n$ from $\mathcal{E}_{\phi_{n}(t_s)}$;\\
						Match $j$ and $n$ setting $\phi_j(t_s)=n$ and $\phi_n(t_s)=j$;\\
						Remove $j$ from $\mathcal{S}^{rx}$
					}
					\Else{ 
						Refuse proposal from $j$;\\
						Remove $n$ from $\mathcal{E}_{j}$; 
					}
				}
			}
		}
		\textbf{Phase IV - Stable matching}
	\end{algorithm}
}
\subsection{V2V Link Selection as a Matching Game}

In order to properly address the fundamentals of this mathematical framework, several definitions must be first done and particularized for the problem at hand:
\begin{definition}
	A matching game is defined by two sets of players ($\mathcal{I}_j(t),\mathcal{J}_i(t)$) and two preference relations $\succ_i$, $\succ_j$, allowing each player $i\in {\mathcal{I}_j(t)}$, $j\in {\mathcal {J}_i(t)}$ to accordingly rank the players in the opposite set. 
\end{definition}

\begin{definition}
	The output of a matching game is a matching function $\bm{\Phi}(t)=\{\phi_{i,j}(t)\}$ that bilaterally assigns players $\bm{\phi}_i(t)\triangleq\{j\in \mathcal{J}_i(t):\phi_{i,j}(t)=1\}$ and $\bm{\phi}_j(t)\triangleq\{i\in \mathcal{I}_j(t):\phi_{i,j}(t)=1\}$ such that $|\bm{\phi}_j(t)|=q_j$ and $|\bm{\phi}_i(t)|=q_i$ are fulfilled. Notice here that $q_i$ and $q_j$ represent the \textit{quota} of the player which, for a one-to-one matching game, $q_i=q_j=1$. 
\end{definition}

\begin{definition}
	A preference $\succ$ is a complete, reflexive and transitive binary relation between the players in $\mathcal{I}_j(t)$ and $\mathcal{J}_i(t)$. Therefore, for any vTx $i$ a preference relation $\succ_i$ is defined over the set of vRx $\mathcal{J}_i(t)$ such that for any two vRx $(m, n)\in \mathcal{J}_i(t)\times \mathcal{J}_i(t)$ with $m\neq n$, and two matchings $\bm{\Phi}(t)$ and  $\bm{\Phi}^\prime(t)$ so that $\bm{\phi}_i(t)=m$ and $\bm{\phi}_i^\prime(t)=n$:
	\begin{equation}
	\left(m,\bm{\Phi}(t)\right)\succ_i\left(n,\bm{\Phi}^\prime(t)\right)\Leftrightarrow U_{vTx}^{i,m}(t)>U_{vTx}^{i,n}(t).\label{eq:pref-tx}		
	\end{equation}
	
	Similarly, for any vRx $j$ a preference relation $\succ_j$ is defined over the set of vTx $\mathcal{I}_j(t)$ such that for any two vTx $(k,l)\in \mathcal{I}_j(t)\times \mathcal{I}_j(t)$ with $k\neq l$, and two matchings $\bm{\Phi}(t)$ and  $\bm{\Phi}^\prime(t)$ so that $\bm{\phi}_j(t)=k$ and $\bm{\phi}_j^\prime(t)=l$:
	\begin{equation}
	\left(k,\bm{\Phi}(t)\right)\succ_j\left(l,\bm{\Phi}^\prime(t)\right)\Leftrightarrow U_{vRx}^{j,k}(t)>U_{vRx}^{j,l}(t), \label{eq:pref-rx}		
	\end{equation}
	where $U_{vTx}^{i,m}(t)$ and $U_{vRx}^{k,j}(t)$ denote the utility of vRx $m$ for vTx $i$ and the utility of vTx $k$ for vRx $j$, correspondingly.
\end{definition}

\begin{definition}
	A matching is not stable if for a given match $\bm{\phi}_i(t)=j$ and $\bm{\phi}_j(t)=i$, a blocking pair $(i^\prime,j^\prime)$ such that $i,i^\prime\in \mathcal{I}_j(t)$ and $j,j^\prime\in \mathcal{J}_i(t)$ satisfying $\bm{\phi}_i(t)\neq j^\prime$, $\bm{\phi}_j(t)\neq i^\prime$ and $j^\prime\succ_i j$, $i^\prime\succ_j i$ exists. That is, if for a given match two players prefer to be matched to each other rather than to their current matched partners. A matching is considered \textit{pairwise stable} if no such blocking pair exists.
\end{definition}

From an algorithmic point of view, Gale-Shapley's Deferred Acceptance algorithm (DA, \cite{galeshapley1962}) provides a polynomial time converging solution for one-to-one canonical matchings i.e.,  those matching games where preferences of players are not influenced by any other player's decisions. To this end DA employs an iterative process which finds a stable mapping from the elements of the set of transmitters in the system at every scheduling period to the elements of the set of feasible receivers. The process relies on the ordering of the preference list that each player on either side compiles over the players from the other set. Let us remark here that DA ensures pairwise stability (as per Definition 4), but is not necessarily optimal for all players in the game. The traditional form of the algorithm is optimal for the initiator of the proposals whereas the stable, suitor-optimal solution may or may not be optimal for their reviewers. Interestingly for the application tackled in this paper, DA does not require a centralized controller as the players involved do not need to observe the actions or preferences of other players.

Unfortunately, the existence of interdependencies between the players' preferences (referred to as \textit{externalities}) makes DA unsuitable as the ranking of preferences lying at its core dynamically changes as the matching evolves. Externalities also pose a great challenge to ensure stability in the matching. 

\subsection{Utility Formulation}

To produce the V2V link allocation that leads to minimum system-wide average delay, participants in the game -- namely, vTxs and vRx in the vehicular scenario at a given scheduling slot -- will determine the utilities perceived towards each other in such a way that this information is captured and used to identify the set of players that offer better delay profiles. The baseline for the formulation of utilities in both vTxs and vRxs will be the $\alpha$-fair utility function \cite{Mo2000} expressed, for $\alpha\ge0$ and $x\in\{vTx,vRx\}$, as 
\begin{equation}
U_x(r_x(t))=\omega_x \frac{r_x(t)^{1-\alpha_x}}{1-\alpha_x}, \label{eq:alphafair}
\end{equation}
where $\alpha=2$ guarantees a weighted minimum proportional delay fairness, and $\omega_x$ allows bringing problem-specific information into the utilities. At this point we recall that $\mathcal{J}_i(t_s)$ and $\mathcal{I}_j(t_s)$ denote the subsets of feasible vRxs for vTx $i$ and feasible vTxs for vRx $j$ at a given scheduling time $t_s \in \mathcal{T}_s$, respectively. With this notation in mind, we define the weighted $\alpha$-fair utility function for vTx $i \in \mathcal{I}(t_s)$ over vRxs $\mathcal{J}_i(t)$ as
\begin{equation}\label{eq:utilityTx}
U_{vTx}^{i,j}\left(t_s\right)\triangleq\hspace{-0.5mm}-\frac{\omega_{vTx}^{i,j}(t_s)}{{r}_{i,j}(t_s,\bm{\Phi}(t_s))},
\end{equation}
where we remark that for notational simplicity we will use $U_{vTx}^{i,j}\left(t_s\right)$ instead of $U_{vTx}^{i,j}\left(t_s,\bm{\Phi}(t_s)\right)$ even though the implicit dependence of the utility on $\bm{\Phi}(t_s)$. Similarly, the utility of vRx $j \in \mathcal{J}(t_s)$ over $I_j(t_s)$ vTxs for a given matching $\bm{\Phi}(t_s)$ will be given by
\begin{equation}\label{eq:utilityRx}
U_{vRx}^{j,i}(t_s)\hspace{-0.5mm}=\hspace{-0.5mm}-\frac{\omega_{vRx}^{j,i}(t_s)}{{r}_{i,j}(t_s,\bm{\Phi}(t_s))},
\end{equation}
so that the system welfare $\mbox{S}(t_s,\bm{\Phi}(t_s))$ to be maximized is
\begin{equation}
\mbox{S}(t_s,\bm{\Phi}(t_s))\hspace{-0.5mm}\triangleq\hspace{-1.5mm} \sum\limits_{\mathcal{I}(t_s)}\sum\limits_{\mathcal{J}_i(t_s)} \hspace{-0.5mm}\phi_{i,j}\hspace{-0.2mm}(t_s)\hspace{-0.2mm}\left(U_{vTx}^{i,j}\hspace{-0.2mm}\left(t_s\right)\hspace{-0.2mm} + \hspace{-0.2mm}U_{vRx}^{j,i}\hspace{-0.2mm}(t_s)\right).
\end{equation}
By including in the expressions of the above utilities --e.g. through weights $\omega_{vTx}^{i,j}(t_s)$ and $\omega_{vRx}^{j,i}(t_s)$-- the traffic influx rate $\rho=\lambda P_s$, the nexus between above utility functions and the fitness in \eqref{eq:mainOptprb} is straightforward. As a result, the above formulated utility functions will reflect the load of the V2V link in terms of the number of transmission slots to serve $\lambda P_s$ bits with rate $r_{i,j}(t_s,\bm{\Phi}(t_s))$. Therefore, the maximization of the system-wide welfare in turn minimizes the fitness in Expression \eqref{eq:fitness}.

We finally define weights $\omega_{vTx}^{i,j}(t_s)$ and $\omega_{vRx}^{j,i}(t_s)$ so that under the same other conditions, vTxs are encouraged to select those vRxs moving along the highway at similar speeds --as that implies links being less prone to misalignment events-- whereas vRxs will choose those vTxs with longer queues in order to alleviate the system. By denoting the relative speed of vTx $i$ and vRx $j$ averaged over the transmission slot $t_s\in\mathcal{T}_s$ as $\overbar{\Delta v}_{i,j}(t_s)$, and the status of queue $i$ at time $t_s$ as $Q_i(t_s,\bm{\Phi}(t_s))$, the proposed weights for the above utility functions are expressed as
\begin{eqnarray}
\omega_{vTx}^{i,j}(t_s)=\rho\left(1+\frac{|\overbar{\Delta v}_{i,j}(t_s)|}{\left|\Delta v\right|_{\max}}\right), \label{eq:alpha-weight-speed}\\
\omega_{vRx}^{j,i}(t_s)=\rho\left(2-\frac{{Q}_{i}(t_s,\bm{\Phi}(t_s))}{Q_{\Theta}}\right), \label{eq:alpha-weight-queue}
\end{eqnarray}
where $i\in\mathcal{I}(t_s)$, $j\in\mathcal{J}(t_s)$, and $|\overbar{\Delta v}|_{\max}$ and $Q_{\Theta}$ represent normalization terms. In the utility \eqref{eq:alpha-weight-queue} we extend the notation in \eqref{eq:queuedynamics} as ${Q}_{i}(t_s,\bm{\Phi}(t_s))$ to denote the queue status at vTx $i$ and time $t_s$ when it is paired to vRx $j$ under matching $\bm{\Phi}(t_s)$.

In practice the need for information exchanges of the current matching state at an instantaneous scale contradicts our overall approach to the problem. Moreover, the formulation of \eqref{eq:utilityTx} and \eqref{eq:utilityRx} reflects that the rate on a link $\ell_{i,j}$ will not only depend on the currently matched vTx, but also on whom the rest of the vTxs are matched to, which unveils the existence of externalities. These externalities in our system are the result of the directionality of mmWave links and the variability of the levels of received interference built upon the beam steering. Unless vRxs are aware of the system-wide current matching, they will not be able to know from which directions interference will arrive and be able to foresee the instantaneous rate of a given mmWave link to cast their preferences. So, with the two-fold aim of reducing instantaneous reporting and of calculating an estimate of ${r}_{i,j}(t_s,\bm{\Phi}(t_s))$, a link exploration and learning procedure will be carried out as explained in the next subsection. 

\subsection{CSI/QSI Information Learning Procedure}\label{subseq:learning}

The evolution of the V2V system dynamics can be described by CSI and QSI as per \eqref{eq:pathloss} and \eqref{eq:queuedynamics}, respectively. As the system evolves, V2V links should be dynamically enforced/released, beamwidths selected and beam steering triggered. However, CSI between devices and QSI at every vTx can only be measured locally and in a distributed fashion. In order to design a CSI/QSI aware long-term RRM policy and yet reduce the exchange of control information, vRxs will collect and process information on measured channel conditions for all transmission slots within a scheduling interval, and exchange it just before the beginning of a new scheduling period. This procedure also holds in the case of vTxs in regards to their QSI estimations.

Upon matching and beam alignment at scheduling slot $t_s-N$, we assume that every vehicle is able to detect and track vTxs and vRxs in its vicinity $\forall t'\in (t_s-N,t_s]$, which can be done by resorting to standard techniques \cite{quek2013small} or more elaborated approaches as in \cite{HPark2015,Kim2013}. During every transmission interval within the scheduling period at hand, random matchings between vehicles in the vicinity of one another are agreed and set over a mmWave control channel deployed in parallel to the main communication beam. The purpose of this control channel is to allow sampling the CSI of every receiver $j\in\mathcal{J}_i(t_s-N)$ in the group when it receives information from a certain transmitter $i\in\mathcal{I}_j(t_s-N)$. This is accomplished by matching at random every single receiver in the system at time $t'$ with any of the transmitters within its neighborhood. From a series of pilot transmissions in this random matching, every receiver $j\in\mathcal{J}(t_s-N)$ infers, based on the received power and by virtue of its knowledge of the relative position and transmit power of the transmitter $i$ to which it is paired and other vehicles nearby, the channel gain $g_{i,j}^c$ as per \eqref{eq:pathloss} and therefrom, an SINR estimation as per \eqref{eq:sinr}. Once this is done, the receiver stores the estimated SINR along with the time instant at which it was produced, and an identifier of the transmitter to whom it was linked to. This process is performed for every receiver in the system and over all transmission slots $t'\in(t_s-N,t_s]$. As a result, all receivers at the end of the scheduling slot have stored a list with entries $\{t',i,\mbox{SINR}_j(t')\}$, with $\mbox{SINR}_j(t')$. 

To learn an estimate $\overbar{r}_{i,j}^{est}(t_s)$ of the average rate that can be expected for the matched pair $(i,j)$ over the next scheduling period, we will inspect the behavior of this rate metric in the recent past (i.e. the previous scheduling period). Yet, instead of treating all samples equally, those more recent in time will be emphasized so as to lessen the impact of older ones \cite{Atkeson1997}. Based on this rationale, $\overbar{r}_{i,j}^{est}(t_s)$ will be computed as
\begin{equation} \label{eq:rateestimated}
\overbar{r}_{i,j}^{est}(t_s)\hspace{-0.8mm}=\hspace{-4mm}\sum_{t'=t_s-N}^{T_s}\hspace{-3.5mm}W(t',i)\left(1\hspace{-0.8mm}-\hspace{-0.5mm}\frac{\tau_{i,j}(t')}{T_t}\right)\hspace{-0.5mm}B\hspace{-0.5mm}\log_2\left(1\hspace{-0.5mm}+\hspace{-0.8mm}\text{SINR}_{j}(t')\hspace{-.5mm}\right),
\end{equation}
where for $\tau_{i,j}(t')$ calculation, equal parameter values to those used for the main communication channel are adopted. Values for weights $W(t',i)$ will be set such that $W(t',i)\neq 0$ if and only if it exists an entry $\{t',i,\mbox{SINR}_j(t')\}$ in the CSI samples acquired by receiver $j$, $W(t',i)\leq W(t'',i)$ if $t'\leq t''$ and imposing  $\sum_{t'\in (t_s-N,t_s]} W(t',i) = 1$ for any $i$ to which receiver $j$ may have been associated to all along the link exploration process in the previous scheduling period. Once rates $\overbar{r}_{i,j}^{est}(t_s)$ have been estimated at receiver $j$ $\forall i\in\mathcal{I}_j(t_s)$, their values are disseminated to its neighboring transmitters, which are now able to infer the average dynamics under which their queue can be flushed. Now that externalities have been removed from the estimated rates of the system, the average queue status at vTx $i$ when communicating to vRx $j$ is not subject to other matched pairs, and can be estimated as $\overbar{Q}_{i,j}(t_s)=P_s/\overbar{r}_{i,j}^{est}(t_s)$. By inserting this estimated CSI/QSI information in Expressions \eqref{eq:utilityTx} and \eqref{eq:utilityRx}, the final utilities to construct the proposed matching game are
\begin{eqnarray}
U_{vTx}^{i,j}\left(t_s\right)\triangleq\hspace{-0.5mm}-\frac{\omega_{vTx}^{i,j}(t_s)}{\overbar{r}_{i,j}^{est}(t_s)}=-\frac{\rho\left(1+\frac{|\overbar{\Delta v}_{i,j}(t_s)|}{\left|\Delta v\right|_{\max}}\right)}{\overbar{r}_{i,j}^{est}(t_s)}, \label{eq:utilityTx-final} \\
U_{vRx}^{j,i}(t_s)\hspace{-0.5mm}=\hspace{-0.5mm}-\frac{\omega_{vRx}^{j,i}(t_s)}{\overbar{r}_{i,j}^{est}(t_s)}=-\frac{\rho\left(2-\frac{\overbar{Q}_{i,j}(t_s)}{Q_{\Theta}}\right)}{\overbar{r}_{i,j}^{est}(t_s)}, \label{eq:utilityRx-final}
\end{eqnarray}
i.e. as a result of the link exploration and learning mechanism, the final utilities for vTxs and vRxs will no longer change during the formation of the game; the V2V mmWave link allocation problem can be cast as a one-to-one canonical matching game and solved by applying the DA algorithm as detailed in Algorithm \ref{algo:DA-v2v}. 

\subsection{Beamwidth Allocation using Swarm Intelligence} \label{subsec:PSO}

Once vTxs and vRxs have been paired by virtue of the matching game explained above and following Fig. \ref{fig:timeScale-flowchart}, an optimal allocation of beamwidths $\bm{\varphi}^{t_x}(t_s)$ and $\bm{\varphi}^{r_x}(t_s)$ for the scheduling slot $t_s\in\mathcal{T}_s$ is performed by using Swarm Intelligence, a family of computational methods capable of efficiently dealing with convex and non-convex hard optimization problems. To this end, Swarm Intelligence relies on systems of interacting agents governed by simple behavioral rules and inter-agent communication mechanisms, such as those observed in certain insects and animal species. In particular we will focus on the so-called Particle Swarm Optimization (PSO \cite{eberhart1995new}), which has been recently utilized to allocate resources in mmWave 5G networks \cite{Perfecto2016,Scott-Hayward2015}. 

Algorithmically the PSO-based beamwidth allocation scheme iteratively updates a $K$-sized swarm of candidate solutions $\{\mathbf{S}\}_{k=1}^K$, which for the problem at hand will be expressed as $\mathbf{S}_k=\bm{\varphi}_k^{t_x}(t_s),\bm{\varphi}_k^{r_x}(t_s)$ with $k\in\{1,\ldots,K\}$ and $\zeta\triangleq|\mathbf{S}_k|$ equal to the number of effective mmWave links established after the matching phase. The algorithm starts by assigning a fixed beamwidth (5\degree) to all beamwidths in $\mathbf{S}_k$, and by setting a velocity vector $\mathbf{V}_{k}=\{V_k^1,\ldots,V_k^{\zeta}\}$ per every candidate solution with inputs initially drawn uniformly at random from the range [5\degree, 45\degree]. The quality of the produced solutions is measured in terms of the average data rate computed over the active mmWave links in the system at time $t_s$. The PSO optimization procedure continues by refining the velocity vector based on its previous value, the best value of $\mathbf{S}_k$ found by the algorithm until the iteration at hand (denoted as $\mathbf{S}_k^\ast=(S_k^{1,\ast},\ldots,S_k^{\zeta,\ast})$), and the global best solution $\mathbf{S}_\triangleright=\{S_{\triangleright}^1,\ldots,S_{\triangleright}^{\zeta}\}$ of the entire swarm as
\begin{equation}
V_k^s \leftarrow \varpi v_k^s + \eta r_{\eta} (S_k^{s,\ast} - S_k^s) + \xi r_{\xi} (S_\triangleright^s - S_k^s),
\end{equation} 
with $s\in\{1,\ldots,\zeta\}$. Once the velocity vector has been updated, the value of every candidate solution $\mathbf{S}_k$ is updated as $\mathbf{S}_k \leftarrow \mathbf{S}_k + \mathbf{V}_k$, from which the best candidates for every particle in the swarm (i.e. $\mathbf{S}_k^\ast$) and the global best candidate $\mathbf{S}_\triangleright$ are recomputed and updated if necessary. Parameters $\varpi$ (inertia), $\eta$ and $\xi$ permit to drive the search behavior of this heuristic, whereas $r_\eta$ and $r_\xi$ are realizations of a uniform random variable with support $[0,1]$. This process is repeated for a fixed number of iterations $\mathcal{I}$. 

\begin{table}[t!]
	\renewcommand{\arraystretch}{1.2}
	\caption{Main Simulation Parameters}\label{tab:SimParameters}
	\centering
	\begin{tabular}{|p{3.95cm}|p{3.95cm}|}
		\hline \textbf{Parameter}&\textbf{Value}\\
		\hline Simulation time& 30000 ms\\
		\hline Transmission slot ($T_t$) & 2 ms\\
		\hline Scheduling slot ($T_s$) & [20, 50, 100, 200, 500] ms\\
		\hline Avg. Vehicle Density& [70, 90, 130, 180] vehicles/km\\	
		\hline Lane Speed & [140, 130, 125, 110, 90, 70] km/h\\
		\hline Car to Truck ratio& 80\% (cars), 20\% (trucks)\\
		\hline vTx/vRx probability& 50\% (vTx), 50\% (vRx)\\
		\hline Coverage radius ($R_c$) & 100 m \\
		\hline Peak transmit/slot time ($T_p/T_s$)&$0.01$\\
		\hline Sector-level beamwidth ($\psi_i^{t_x}$,$\psi_i^{r_x}$)&45\degree\\
		\hline Carrier frequency & 60 GHz\\
		\hline Bandwidth ($B$)& 2.16 GHz\\
		\hline Noise Power Spectral density ($N_0$)&-174 dBm/Hz\\
		\hline vTx transmit power ($p_{i}$) & 15 dBm\\
		\hline Packet size ($P_s$) & [3200, 10$^6$, 2097144, 10$^7$] bits\\
		\hline Mean traffic arrival rate ($\lambda$) &[1/2, 1/6, 1/20, 1/60] packets/ms\\
		\hline
	\end{tabular}
\end{table}

\section{Simulation Setup and Results} \label{sec:results}

In order to assess the performance of the proposed scheme comprehensive computer experiments have been performed over a 500 meter-long highway segment with 6 lanes of 3m width each. Vehicles are assumed to move in the same direction at constant speeds of --leftmost to rightmost lane-- 140, 130, 125, 110, 90, and 70 km/h. Vehicles are either cars (80\%) or trucks  (20\%), with cars drawn uniformly at random from a set of 5 different models, each with varying lengths and widths. Four scenarios with traffic densities of $\{70, 90, 130, 180\}$ vehicles/km will be considered in the experiments and, hereafter, referred to as \texttt{LOW}, \texttt{MID}, \texttt{HIGH} and \texttt{ULTRA}. In order to fix the vehicle density at every scenario, vehicles leaving the segment will trigger the process for new ones to join in, which will be done by prioritizing least crowded lanes, and by guaranteeing a minimum distance to the preceded vehicle. Upon their entrance to the road, vehicles will be declared as transmitters (vTx) or receivers (vRx) with equal probability. Disregarding the role of those vehicles leaving the system, the new ones will be endorsed as vTx or vRx indistinctly.

According to Table \ref{tab:SimParameters}, the highway road scenario has been simulated for a total time of 30000 ms, with transmission intervals of $T_t=2$ ms and scheduling intervals $T_s\in\{20, 50, 100, 200, 500\}$ ms. To assess the impact of queue dynamics under different configurations several packet arrival rates and sizes\footnote{Note that packet sizes of $P_s=3200$ and $P_s=2097144$ bits are in line with the specifications for the DSRC safety messages length \cite{Kenney:11} and the 802.11ad maximum payload \cite{perahia2011gigabit}, respectively.} are considered. 
\begin{figure*}[ht]
\vspace{-.6cm}
	\captionsetup[subfigure]{aboveskip=1pt}
	\begin{tabular}{@{}c@{}}
		\subfloat[]{
			\includegraphics[width=.97\columnwidth]{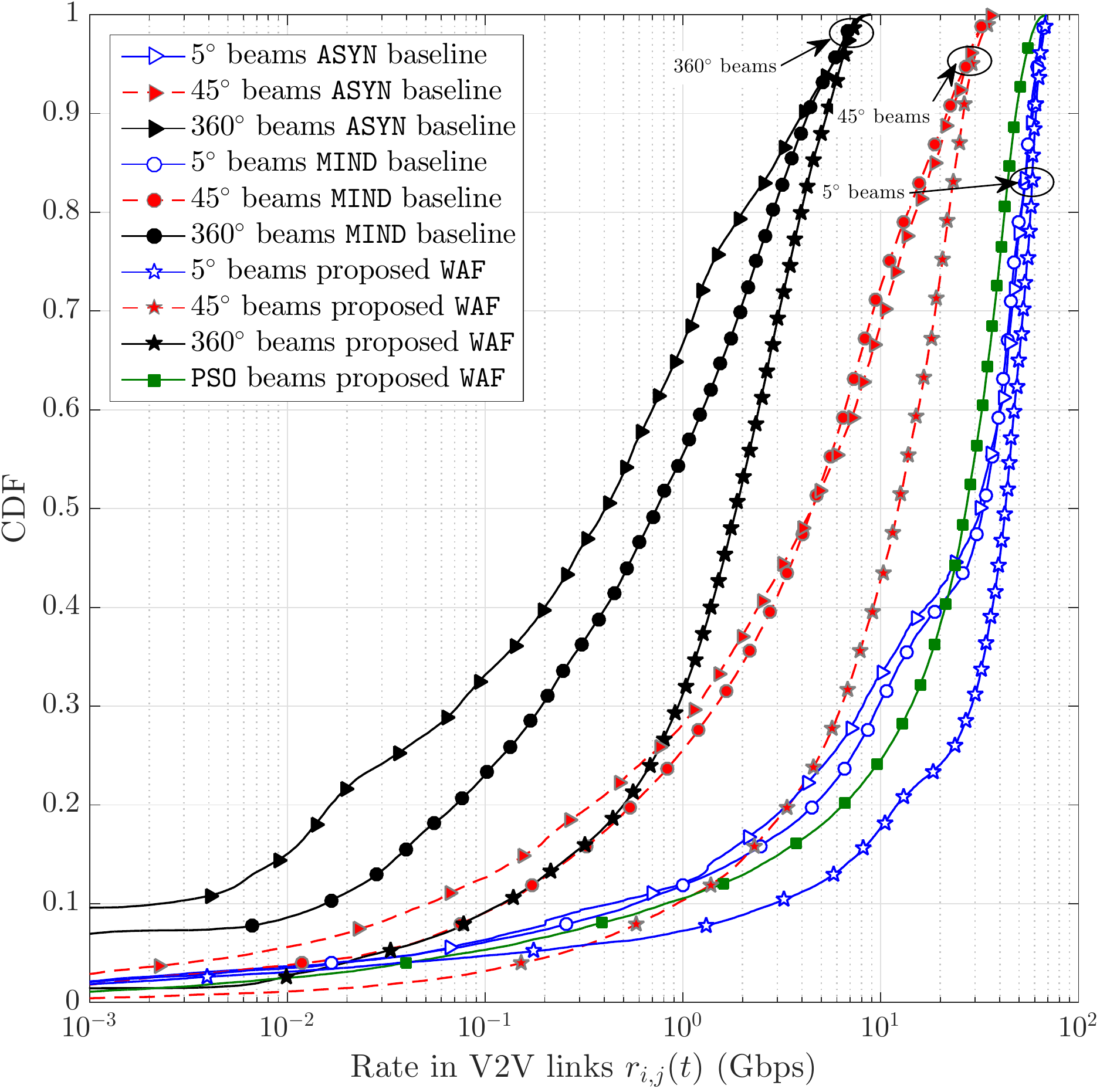}\label{fig:MultiBeamW-rateCDF-shortpacket}}
		\subfloat[]{
			\includegraphics[width=.965\columnwidth]{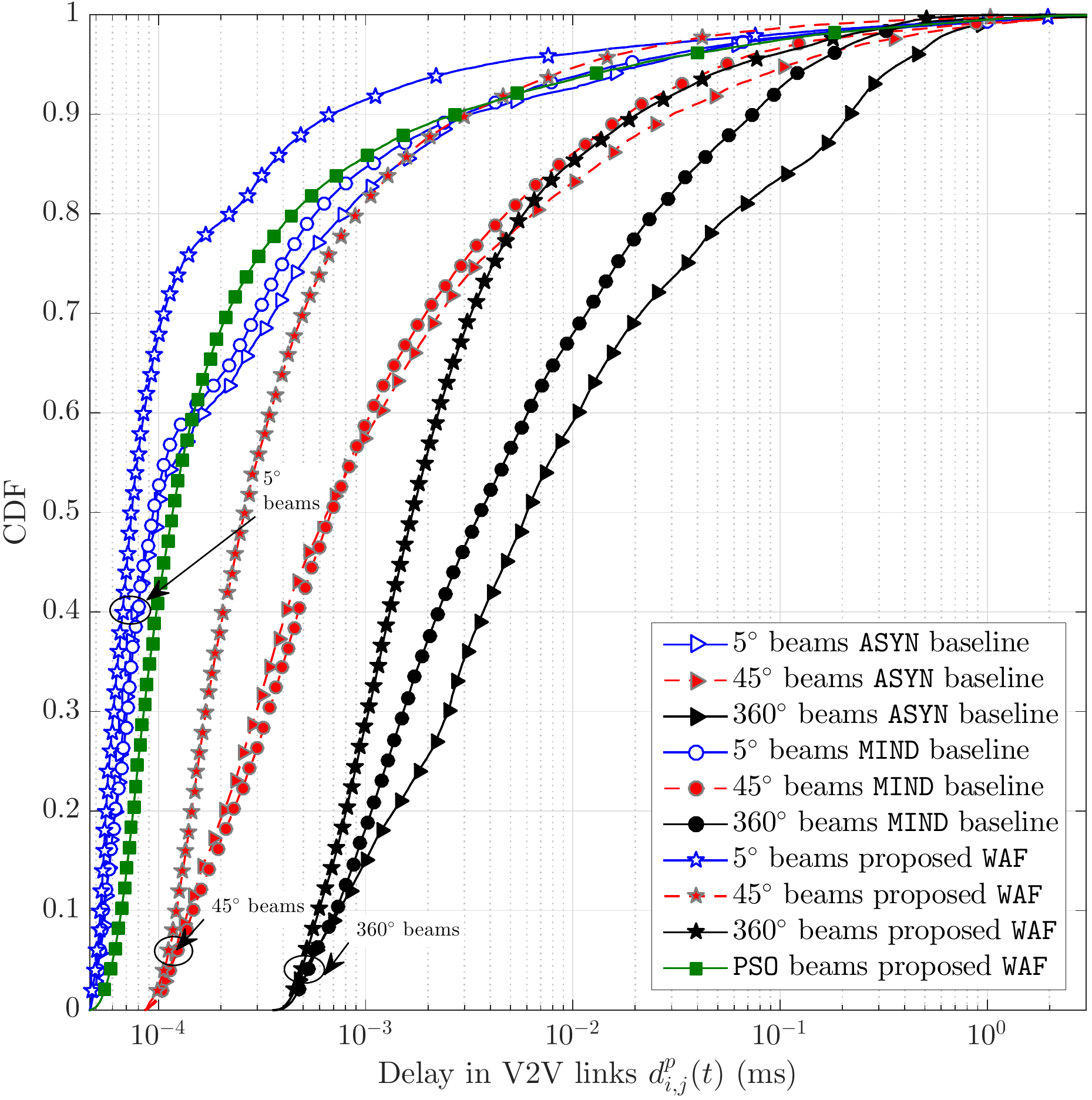}\label{fig:MultiBeamW-delayCDF-shortpacket}}
	\end{tabular}
	\begin{tabular}{@{}c@{}}
		\subfloat[]{
			\includegraphics[width=.965\columnwidth]{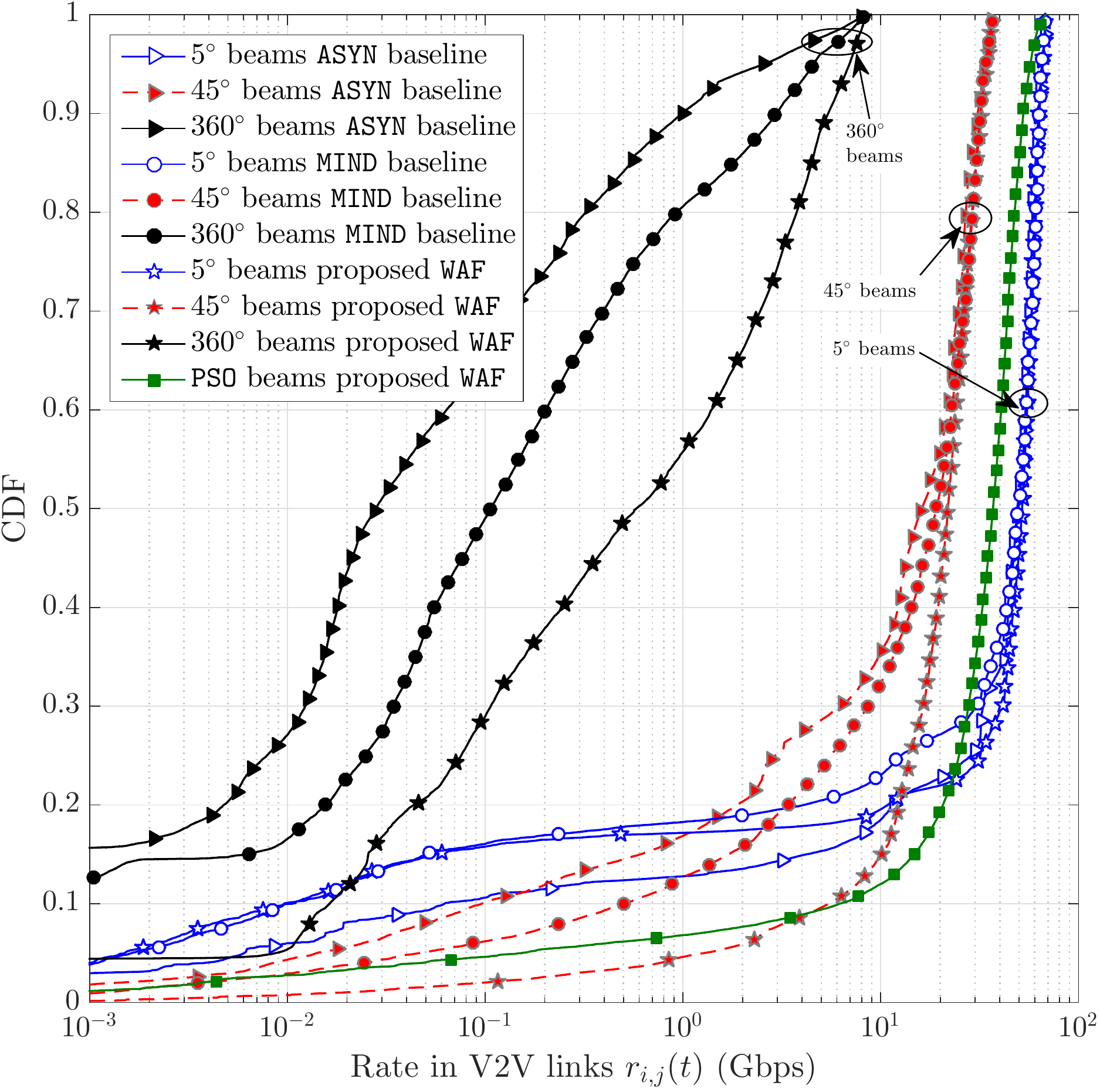}\label{fig:MultiBeamW-rateCDF-longpacket}}
		\subfloat[]{
			\includegraphics[width=.965\columnwidth]{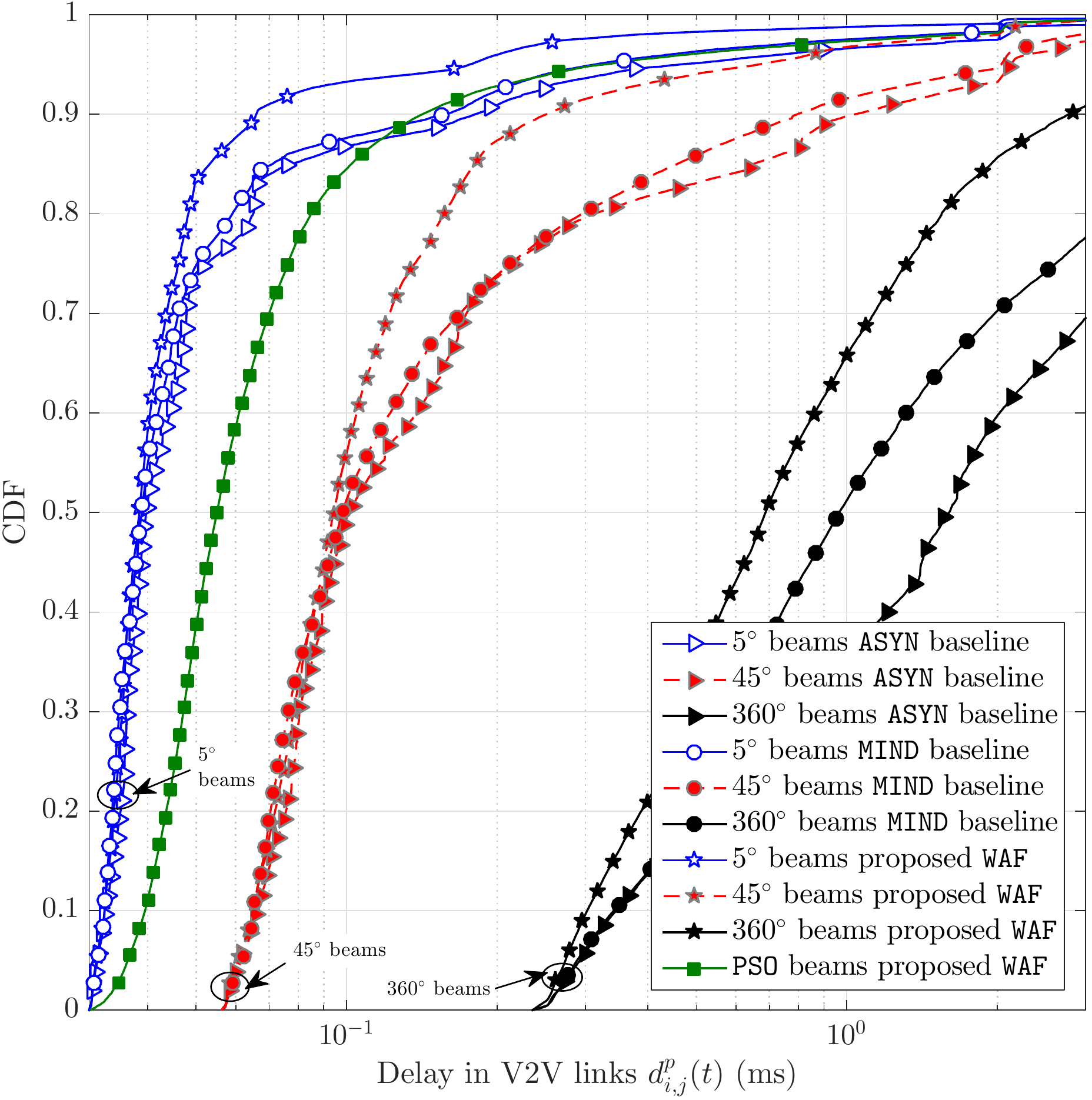}\label{fig:MultiBeamW-delayCDF-longpacket}}
	\end{tabular}		
	\caption{Rate and delay CDFs of the baseline and proposed approaches in $\texttt{ULTRA}$ density scenario for different beamwidths:  (a) and (b) for $P_s=3200$ bits and traffic arrival rate $\lambda=1/T_t$ packets/s; (c) and (d) for $P_s=2097144$ bits and traffic arrival rate $\lambda=1/10\cdot T_t$ packets/s.}
\label{fig:CDFRateDelay}
\end{figure*}

As shown in Fig. \ref{fig:timeScale-flowchart}, two variants of our V2V allocation method will be considered for discussion:
\begin{itemize}
\item Fixed-beamwidth weighted $\alpha$-fair matching (\texttt{WAF}), in which the aforementioned deferred acceptance matching algorithm is applied every $T_s$ ms considering the learned utilities as per \eqref{eq:utilityTx-final} and \eqref{eq:utilityRx-final}. In this case transmit and receive beamwidths of the mmWave channels are kept equal for every link. In particular beamwidths of 5$^\circ$, 45$^\circ$ and 360$^\circ$ will be considered.

\item PSO weighted $\alpha$-fair matching (\texttt{PSO}), similar to the scheme above but incorporating the beamwidth optimization phase explained in Section \ref{subsec:PSO}. As detailed therein, this optimization phase is based on the interplay between alignment delay and the throughput in mmWave communications. In all cases the PSO approach uses $K=30$ particles, $\varpi=0.5$, $\eta=\xi=1.5$ and $\mathcal{I}=50$ iterations. As opposed to the \texttt{WAF} approach, this scheme requires a central controller (e.g. a RSU) to coordinate the selection of transmission and reception beamwidths for each vehicle pair. Nevertheless it is of interest to explore this solution to address more realistic scenarios subject to more frequent misalignment events between pairs.
\end{itemize}

Simulation results for the above approaches will be compared to those produced by 2 different baseline schemes contributed in \cite{HeathSurvey2016}, namely:
\begin{itemize}
\item Minimum-distance based pairing (\texttt{MIND}), by which every vTx in the system at a given scheduling slot tries to pair with its closest vRx that has not been paired yet. Pairing is conducted in increasing order of the distance from the vehicle to the beginning of the highway segment. Pairing is renewed as in our framework, i.e. every $T_s$ ms.
\item Asynchronous long-term pairing (\texttt{ASYN}), by which a restrictive distance-based pairing is triggered every time a new vehicle enters the highway segment. Specifically, two vehicles are paired if 1) they are eligible for pairing, i.e. still single and located within the first 20 meters of the highway segment; and 2) they are in the same or adjacent lanes. Once vehicles are associated, the pair remains unchanged until one of them leaves the segment, forcing the other vehicle to be unmatched while on track. 
\end{itemize}
\begin{figure*}[t]
\vspace{-.6cm}
	\begin{tabular}{@{}c@{}}
		\subfloat[]{
			\includegraphics[width=0.95\columnwidth]{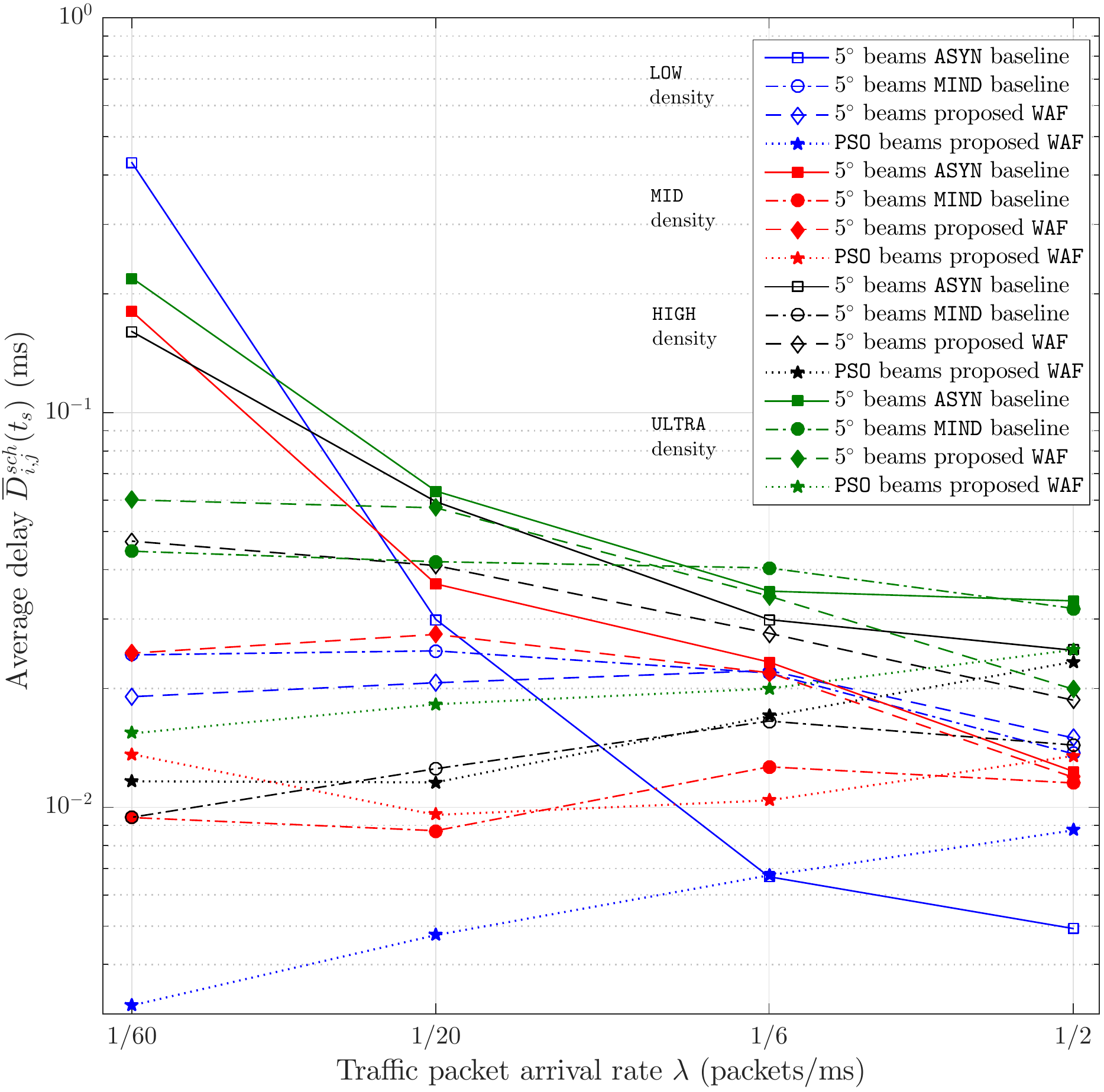}\label{short-delay}}\hfill
		\subfloat[]{
			\includegraphics[width=0.95\columnwidth]{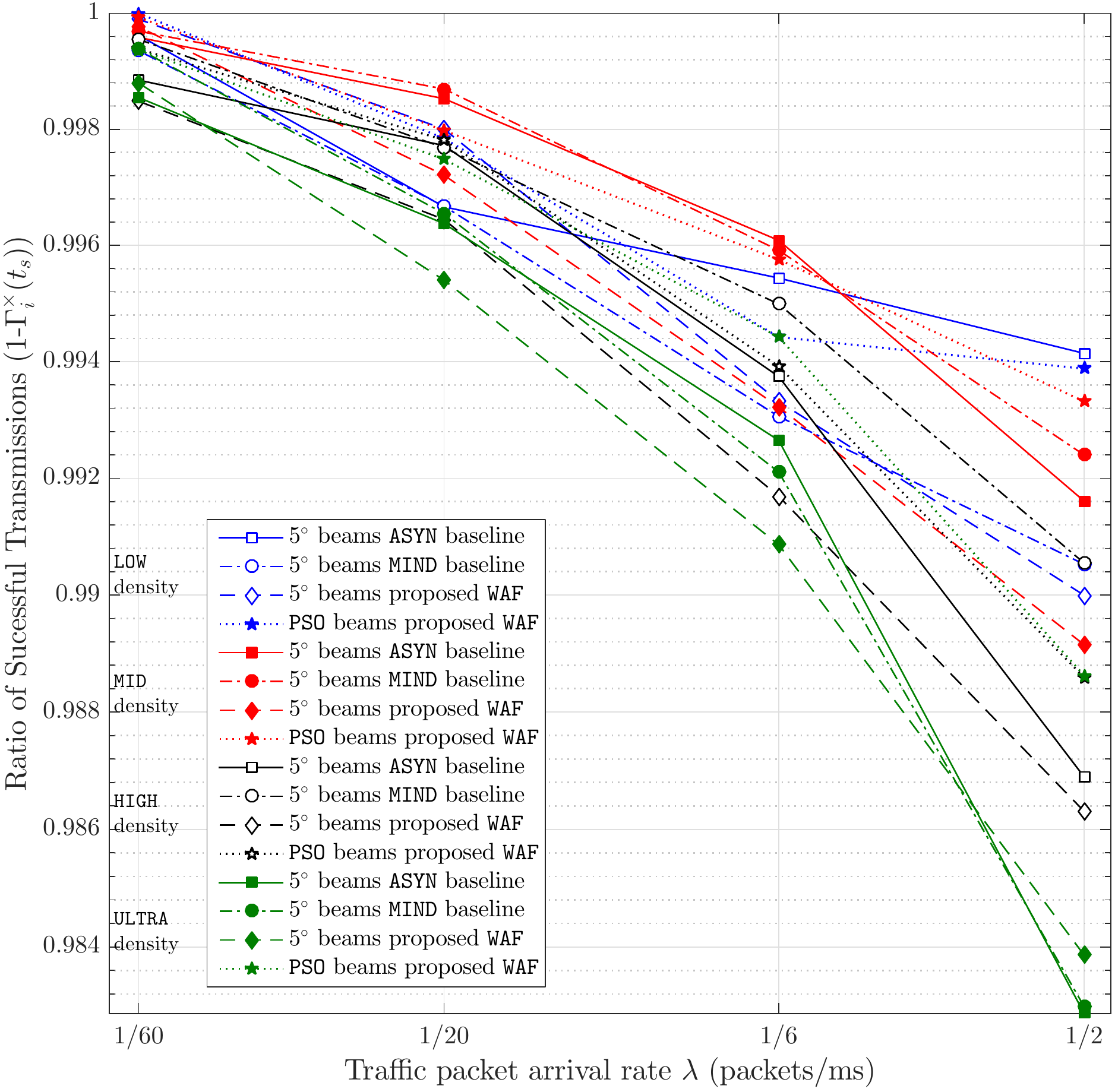}\label{short-success}}
	\end{tabular}\label{fig:DelayVsSuccess-multidensity-ShortPacket}		
	\begin{tabular}{@{}c@{}}
		\subfloat[]{
			\includegraphics[width=0.95\columnwidth]{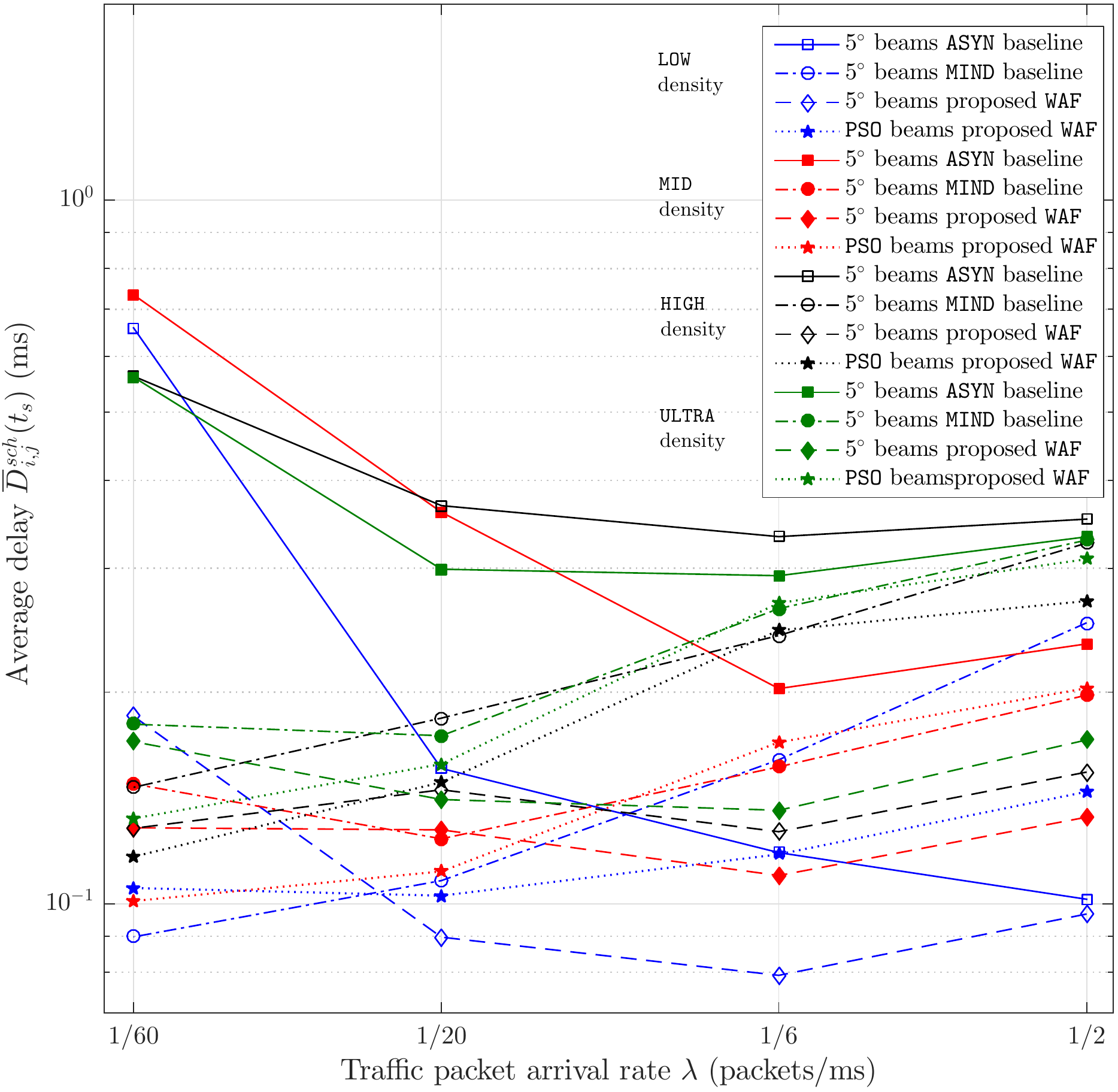}\label{long-delay}}\hfill
		\subfloat[]{
			\includegraphics[width=0.95\columnwidth]{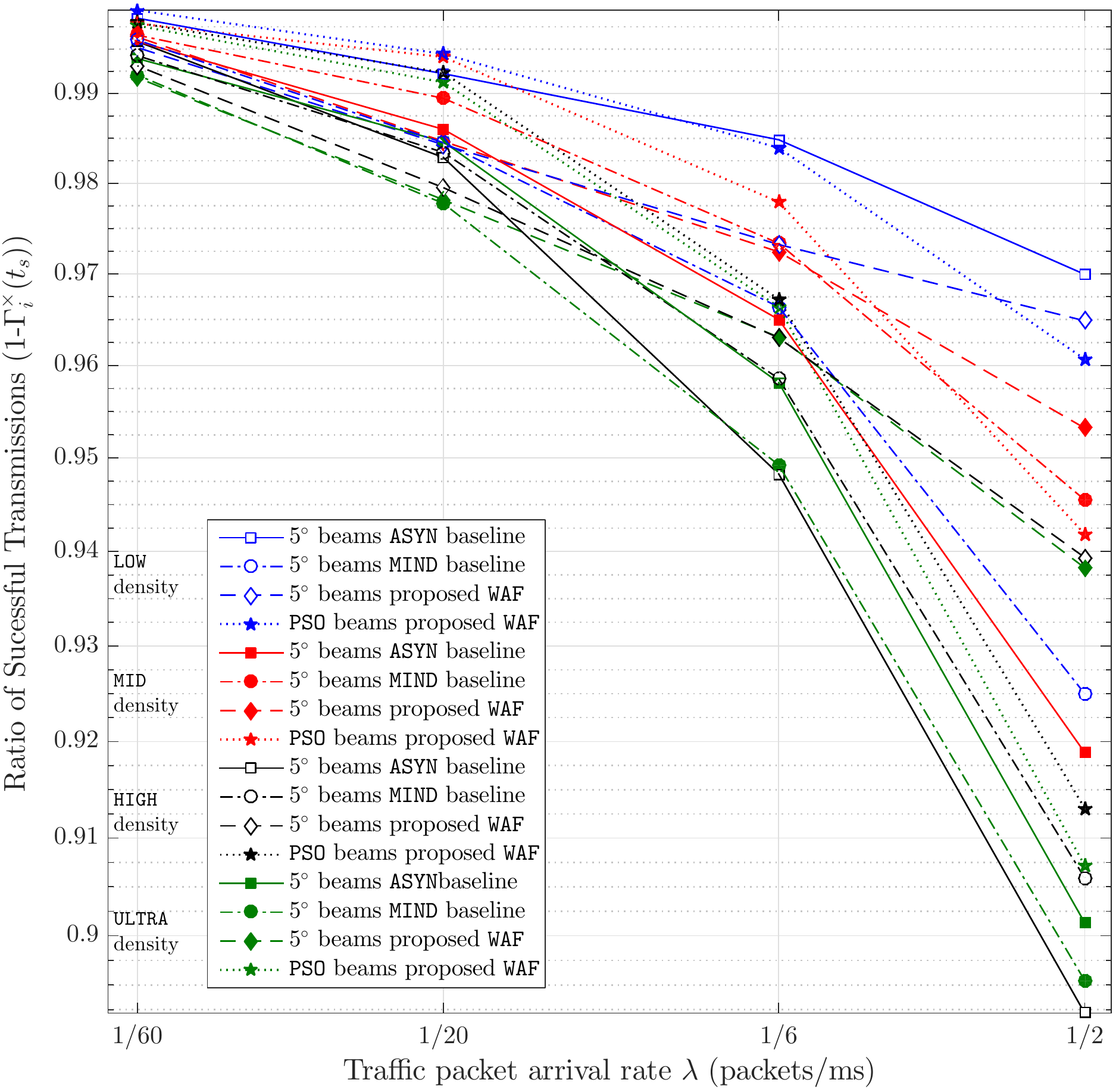}\label{long-success}}
	\end{tabular}\label{fig:DelayVsSuccess-multidensity-longPacket}	
\caption{Interplay between delay and transmission success under different vehicle densities and traffic arrival rate configurations: (a) delay and (b) successful transmissions for the short packet case, $P_s=3200$ bits; (c) delay and (d) successful transmissions for the long packet case, $P_s=2097144$ bits.}	
\label{fig:DelayVsSuccess-multidensity}	
\end{figure*}

In all the above methods matching and pairing strategies will be subject to coverage constraints arriving from $R_c$. Thus, unpaired vTx/vRx might stem from asymmetries in the number of vTx and vRx at a given time slot. Moreover, coverage constraints might yield singleton vTxs and vRxs due to an infeasible association between remaining candidates.

\subsection{Discussion}\label{subsec:results}

Before proceeding further with the analysis let us remark here that a proper interpretation of the obtained results should simultaneously consider delay and reliability statistics. The reason lies in the stringent packet dropping policy adopted in this work, which deducts from the delay calculation as per \eqref{eq:delay_scheduling_comp} packets not fulfilling a delay below $D_{\max}^\lambda$ set for simulations such that $D_{\max}^\lambda=1/\lambda$. In this context, packets in queues with associated transmission rates matching or exceeding the traffic influx rate will contribute to delay statistics, whereas those in  queues with slower rates will be more likely to be dropped. Therefore, as the number of packets successfully transmitted within $D_{\max}^\lambda$ decreases so does the number of transmissions contributing to queue average delay calculations that will be, in any case, upper bounded by $D_{\max}^\lambda$. Another indicator that should be considered when evaluating the goodness of all pairing approaches in this benchmark is the number of effectively matched vehicles. In this regard, it can be expected that the \texttt{ASYN} method fails to pair as many vehicles as the rest of the schemes, with notable differences that will be quantified next. Finally, we restrict the discussion to some representative $(P_s,\lambda)$ combinations: $(3200\text{ bits}, 1/2\text{ packets/ms})$, characterizing intensive short-length messages transmissions that are common in safety related V2X communications scenarios; and $(2097144\text{ bits}, 1/20\text{ packets/ms})$ and $(2097144\text{ bits}, 1/60\text{ packets/ms})$, which model long packets arriving at a lower rate as for infotainment applications. 

In the remaining of this subsection we will concentrate our discussion towards different purposes. To begin with, the effect of the beamwidth selection will be analyzed through Fig. \ref{fig:CDFRateDelay}. Therein the rate and delay per packet\footnote{For all methods with fixed beamwidths the beam alignment delay is given in \eqref{eq:tau} and implicitly included in the delay computations.} Cumulative Density Functions (CDF) are plotted under \texttt{ASYN}, \texttt{MIND}, and \texttt{WAF} methods for fixed and \texttt{PSO} beamwidths in \texttt{ULTRA} scenario. If we have a closer look to the rate CDF from Fig. \ref{fig:CDFRateDelay}(a) and compare it with the CDF from Fig. \ref{fig:CDFRateDelay}(c) the latter shows much longer tails. Serving longer packets even with lower traffic arrival rates implies an increased system utilization --defined as the ratio of slots where vTxs are engaged in transmission-- and consequently a higher interference which degrades the measured SINR and the link rate. Therefore, the increased delays in Fig. \ref{fig:CDFRateDelay}(d) as compared to those of Fig. \ref{fig:CDFRateDelay}(b) cannot be merely attributed to the increased serving time expected for longer packets. It can be concluded from these plots that narrow beams and PSO-optimized beams render better delay and rate results than any other considered beamwidths. This outperforming behavior holds not only for the plots shown here, but also for other simulated cases not shown in the paper for the sake of brevity. Based on this rationale, from this point onwards discussions will be restricted to the methods with narrow beams and the \texttt{PSO} method.
\begin{table*}[t!]
	\begin{minipage}{.48\textwidth}
		\caption{Percentage of scheduling periods jointly fulfilling $\overbar{D}_{i,j}^{sch}(t_s)$ and $\Gamma_{i}^\times(t_s)$ upper-bounds in \texttt{ULTRA}, $P_s=3200$ bits, $\lambda=1/T_t$ packets/s.}
		\label{tab:delay_drop_bounds_1}
		\centering
		\resizebox{1\columnwidth}{!}{
			\begin{tabular}{|P{1cm}|P{1cm}|P{1.5cm}|P{0.9cm}|P{0.9cm}|P{0.9cm}|P{0.9cm}|P{0.9cm}|}
				\hline 
				\multicolumn{3}{|c|}{\multirow{2}{*}{\texttt{ULTRA}, $3200$, $1/2$ms}} & \multicolumn{5}{c|}{Upper bound for $\Gamma_{i}^\times(t_s)$} \\ 
				\cline{4-8}
				\multicolumn{3}{|c|}{} & $10$\% & $1$\% & $0.1$\% & $0.01$\% & $0.001$\% \\ 
				\cline{1-8} 
				\parbox[t]{4mm}{\multirow{20}{*}{\rotatebox[origin=c]{90}{Upper bound for $\overbar{D}_{sch}(t_s)$}}} & \multirow{4}{*}{\specialcell{$0.1$\\ms}} & \texttt{ASYN} & 94.67&	44.67&	29.67&	28.67&	28.67\\
				\cline{3-8} &  & \texttt{MIND} & 96.32&	35.79&	19.73&	18.06&	18.06\\
				\cline{3-8} & & \texttt{WAF} & 100.00&	38.80&	21.74&	20.07&	20.07\\
				\cline{3-8} &  & \texttt{PSO} & 97.32&	58.19&	37.46&	35.79&	35.79\\
				\cline{2-8} & \multirow{4}{*}{\specialcell{$0.075$\\ms}} & \texttt{ASYN} & 89.00&	44.67&	29.67&	28.67&	28.67\\
				\cline{3-8} &  & \texttt{MIND} & 90.64&	35.79&	19.73&	18.06&	18.06\\
				\cline{3-8} & & \texttt{WAF} & 97.99&	38.80&	21.74&	20.07&	20.07\\
				\cline{3-8} &  & \texttt{PSO} & 95.65&	58.19&	37.46&	35.79&	35.79\\
				\cline{2-8} & \multirow{4}{*}{\specialcell{$0.05$\\ms}} & \texttt{ASYN} & 76.67&	44.00&	29.33&	28.33&	28.33\\
				\cline{3-8} &  & \texttt{MIND} & 79.26&	35.79&	19.73&	18.06&	18.06\\
				\cline{3-8} & & \texttt{WAF} & 89.97&	38.46&	21.41&	19.73&	19.73\\
				\cline{3-8} &  & \texttt{PSO} & 85.95&	56.86&	36.79&	35.12&	35.12\\
				\cline{2-8} & \multirow{4}{*}{\specialcell{$0.025$\\ms}} & \texttt{ASYN} & 54.00&	39.33&	28.33&	27.33&	27.33\\
				\cline{3-8} &  & \texttt{MIND} & 51.51& 31.77&	18.73&	17.39&	17.39\\
				\cline{3-8} & & \texttt{WAF} & 70.57&	33.78&	19.40&	18.06&	18.06\\
				\cline{3-8} &  & \texttt{PSO} & 61.87&	47.83&	32.44&	31.10&	31.10\\
				\cline{2-8} & \multirow{4}{*}{\specialcell{$0.01$\\ms}} & \texttt{ASYN} & 31.33&	26.67&	22.33&	22.33&	22.33\\
				\cline{3-8} &  & \texttt{MIND} & 30.44&	23.41&	16.39&	15.05&	15.05\\
				\cline{3-8} & & \texttt{WAF} & 39.80&	21.07&	15.39&	14.05&	14.05\\
				\cline{3-8} &  & \texttt{PSO} & 36.12&	31.44&	22.74&	22.07&	22.07\\
				\hline 
			\end{tabular}
		}
	\end{minipage}\hfill
	\begin{minipage}{.48\textwidth}
		\caption{Percentage of scheduling periods jointly fulfilling $\overbar{D}_{i,j}^{sch}(t_s)$ and $\Gamma_{i}^\times(t_s)$ upper bounds in \texttt{ULTRA}, $P_s=2097144$ bits, $\lambda=1/30\cdot T_t$ packets/s.}
		\label{tab:delay_drop_bounds_2}
		\centering
		\resizebox{1\columnwidth}{!}{
			\begin{tabular}{|P{1cm}|P{1cm}|P{1.5cm}|P{0.9cm}|P{0.9cm}|P{0.9cm}|P{0.9cm}|P{0.9cm}|}
				\hline 
				\multicolumn{3}{|c|}{\multirow{2}{*}{\texttt{ULTRA}, $2097144$, $1/60$ms}} &\multicolumn{5}{c|}{Upper bound for $\Gamma_{i}^\times(t_s)$} \\ 
				\cline{4-8}
				\multicolumn{3}{|c|}{} & $20$\% & $15$\% & $10$\% & $1$\% & $0.1$\% \\ 
				\cline{1-8} 
				\parbox[t]{4mm}{\multirow{20}{*}{\rotatebox[origin=c]{90}{Upper bound for $\overbar{D}_{sch}(t_s)$}}} 
				& \multirow{4}{*}{\specialcell{$0.5$\\ms}} & \texttt{ASYN} & 71.00&	71.00&	71.00&	62.33&	62.00\\
				\cline{3-8} &  & \texttt{MIND} & 89.63&	89.63&	89.63&	64.88&	61.87\\
				\cline{3-8} & & \texttt{WAF} & 89.63&	89.63&	89.63&	64.55&	62.88\\
				\cline{3-8} &  & \texttt{PSO} & 96.66&	96.66&	96.66&	87.63&	86.29\\
				\cline{2-8} & \multirow{4}{*}{\specialcell{$0.2$\\ms}} & \texttt{ASYN} & 59.67&	59.67&	59.67&	54.00&	53.67\\
				\cline{3-8} &  & \texttt{MIND} & 79.93&	79.93&	79.93&	59.20&	57.53\\
				\cline{3-8} & & \texttt{WAF} & 80.60&	80.60&	80.60&	57.86&	56.52\\
				\cline{3-8} &  & \texttt{PSO} & 90.64&	90.64&	90.64&	83.28&	81.94\\
				\cline{2-8} & \multirow{4}{*}{\specialcell{$0.1$\\ms}} & \texttt{ASYN} & 44.33&	44.33&	44.33&	40.67&	40.33\\
				\cline{3-8} &  & \texttt{MIND} & 70.23&	70.23&	70.23&	51.51&	50.17\\
				\cline{3-8} & & \texttt{WAF} & 76.25&	76.25&	76.25&	54.52&	53.18\\
				\cline{3-8} &  & \texttt{PSO} & 63.55&	63.55&	63.55&	58.53&	57.19\\
				\cline{2-8} & \multirow{4}{*}{\specialcell{$0.075$\\ms}} & \texttt{ASYN} & 40.33&	40.33&	40.33&	36.67&	36.33\\
				\cline{3-8} &  & \texttt{MIND} & 59.53&	59.53&	59.53&	42.81&	41.47\\
				\cline{3-8} & & \texttt{WAF} & 70.90&	70.90&	70.90&	49.83&	48.83\\
				\cline{3-8} &  & \texttt{PSO} & 40.13&	40.13&	40.13&	37.46&	36.79\\
				\cline{2-8} & \multirow{4}{*}{\specialcell{$0.05$\\ms}} & \texttt{ASYN} & 23.00&	23.00&	23.00&	20.33&	20.33\\
				\cline{3-8} &  & \texttt{MIND} & 37.12&	37.12&	37.12&	26.09&	25.08\\
				\cline{3-8} & & \texttt{WAF} & 49.16&	49.16&	49.16&	33.11&	32.44\\
				\cline{3-8} &  & \texttt{PSO} & 00.00&	00.00&	00.00&	00.00&	00.00\\
				\hline 
			\end{tabular}
		}
	\end{minipage}\hfill
\end{table*} 
\begin{figure*}[ht]
	\begin{tabular}{ccc}
		\subfloat[]{\includegraphics[width=.96\columnwidth]{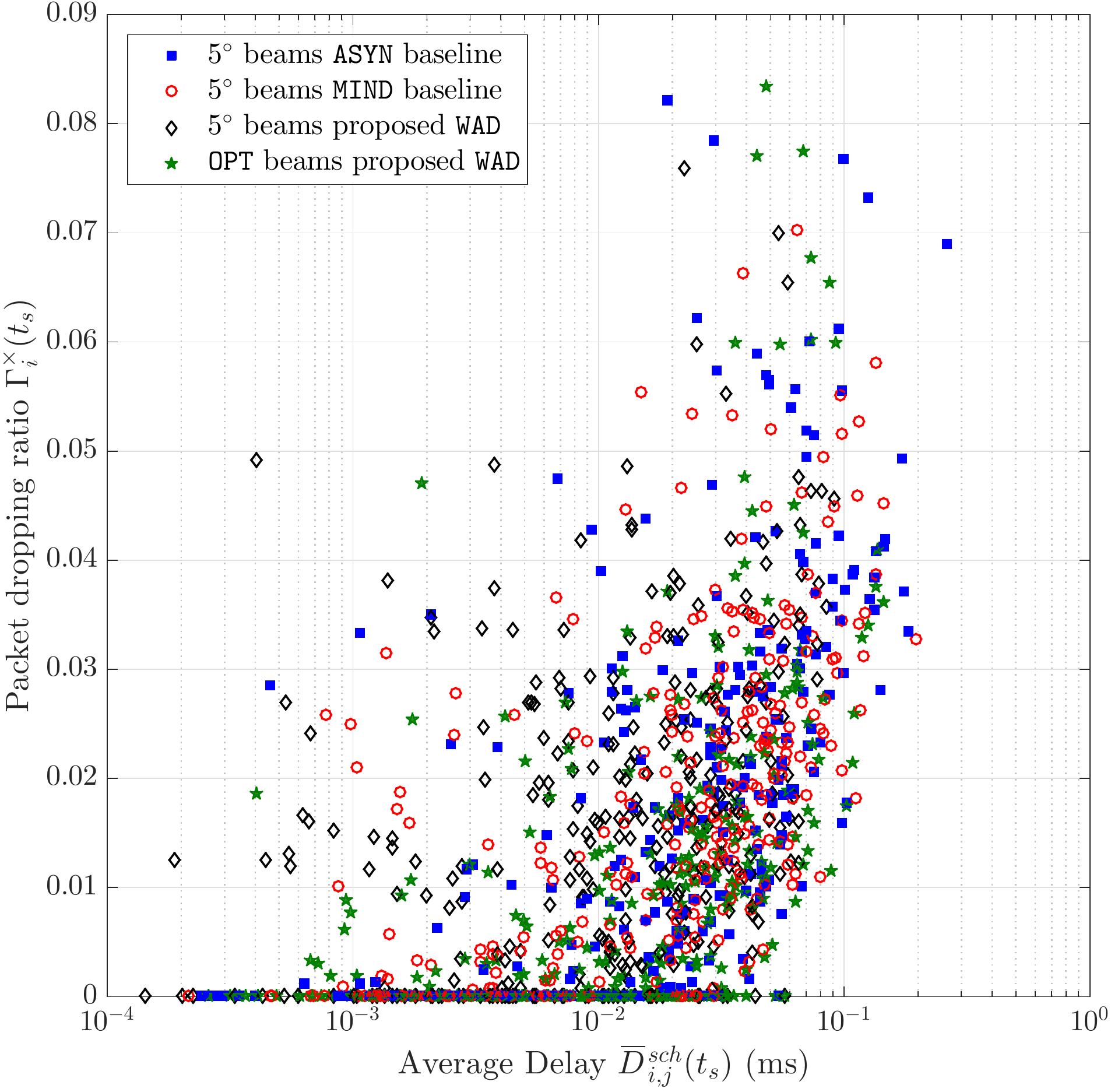}\label{fig:fig4scatter_1}}&&			
		\subfloat[]{\includegraphics[width=.96\columnwidth]{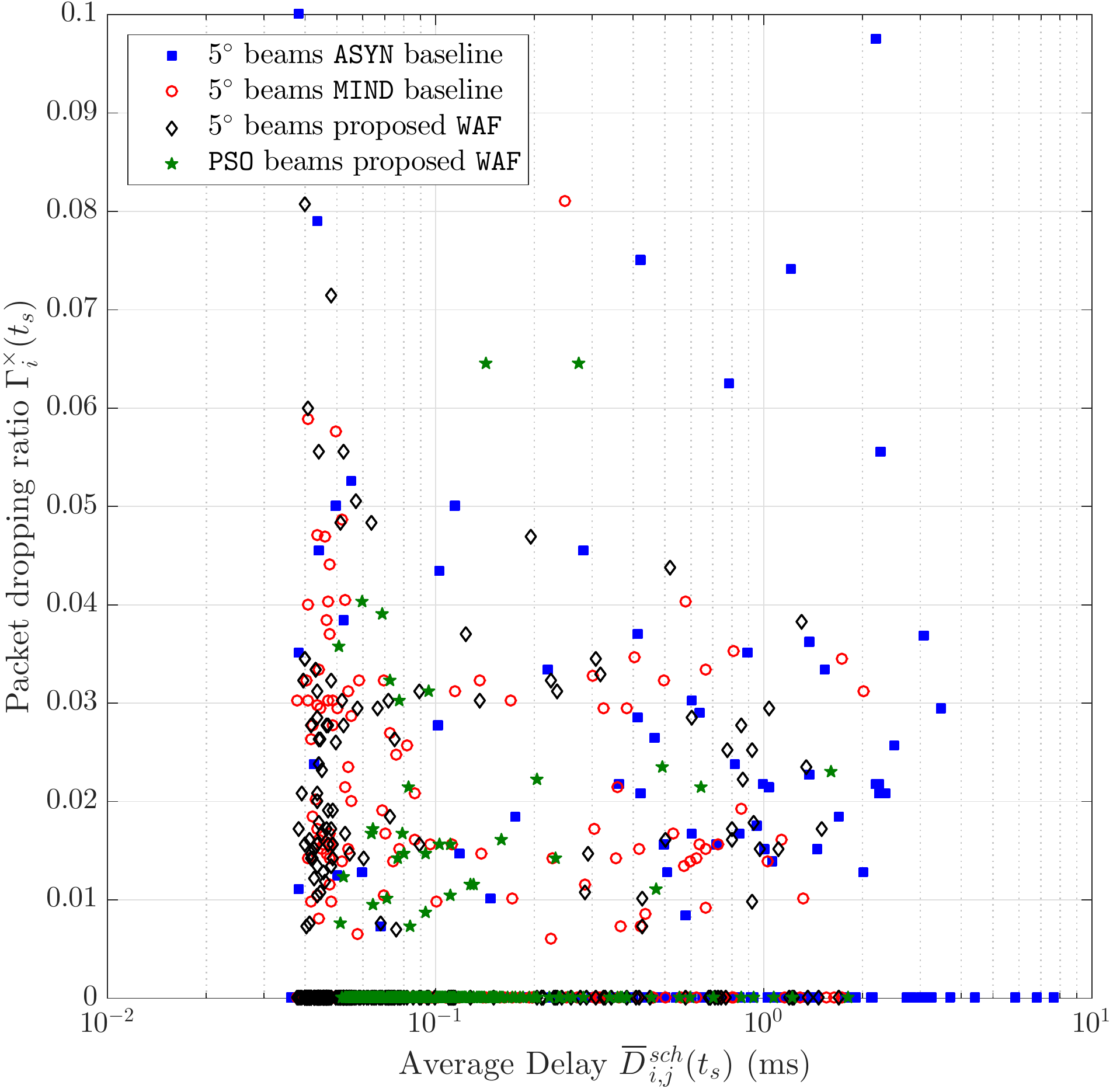}\label{fig:fig4scatter_2}}
	\end{tabular}
	\caption{Scatter plot of $\overbar{D}_{sch}(t_s)$ and $\Gamma_{i}^\times(t_s)$ performance in $\texttt{ULTRA}$ density scenario: (6a) $P_s$=3200 bits, traffic arrival rate $\lambda=1/T_t$ packets/s; (6b) $P_s=2097144$ bits, traffic arrival rate $\lambda=1/30\cdot T_t$ packets/s.}
	\label{fig:fig4scatter}
\end{figure*}
The discussion follows through Fig. \ref{fig:DelayVsSuccess-multidensity}, which further exposes the combined effect of increasing traffic arrival rates on the average delay (Fig. \ref{fig:DelayVsSuccess-multidensity}(a) and Fig. \ref{fig:DelayVsSuccess-multidensity}(c)) and on the average ratio of successful transmissions (Fig. \ref{fig:DelayVsSuccess-multidensity}(b) and Fig. \ref{fig:DelayVsSuccess-multidensity}(d)) under \texttt{LOW}, \texttt{MID}, \texttt{HIGH}, and \texttt{ULTRA} vehicle density scenarios. The effect of the queue dropping policy on the delay is evinced in these plots; while, as expected, the ratio of successful transmissions severely decreases as the traffic arrival rate becomes more demanding, the average delay decreases disregarding the utilized scheme. In other words, those cases where the degradation of the average delay with increasing values of $\lambda$ is not sharp reflect a better resiliency of the system with respect to the traffic arrival rate. However, it must be interpreted along with the ratio of successful transmissions of the method at hand. This being said, from the plots in Fig. \ref{fig:DelayVsSuccess-multidensity} it can be observed that our proposed schemes feature the lowest dropping ratio and the most notable delay resiliency for the most demanding setting (\texttt{ULTRA} vehicle density, $P_s=3200$). As the density becomes lower, performance gaps become smaller, to the point where \texttt{ASYN} offers the highest success ratio for the \texttt{LOW} density scenario. However, the number of vTx paired by the \texttt{ASYN} approach is around $25$\% of the overall number of vTx, whereas for the remaining schemes this number is around $60$\%, increasing to levels above $90$\% in scenarios with higher density. When turning to longer sized packets, dropping ratios increase significantly (more than one order of magnitude).

We now focus the discussion on Table \ref{tab:delay_drop_bounds_1} which shows, for the \texttt{ULTRA} vehicle density case, $N=50$, $\lambda=1/T_t$ and $P_s=3200$, the ratio of scheduling periods $t_s\in\mathcal{T}_s$ over the entire simulation with an average delay $\overbar{D}_{i,j}^{sch}(t_s)$ as per \eqref{eq:fitness} and a packet dropping ratio $\Gamma_{i}^\times(t_s)$ as per \eqref{eq:packet_drop_ratio_sch} --averaged over $t\in[t_s,t_s+NT_t)$-- below different upper bounds. For a better understanding of this table, Fig. \ref{fig:fig4scatter}(a) depicts, for every scheme in the benchmark, the average delay and packet dropping ratio of every scheduling period as a scatter plot. The statistics shown in Table \ref{tab:delay_drop_bounds_1} correspond to the number of points (i.e. scheduling periods) for each matching method that jointly meet upper constraints in both axes. For instance, we can observe that $44.67$\% of the total scheduling periods simulated for the \texttt{ASYN} scheme and the \texttt{ULTRA} dense scenario achieve an average delay below $0.1$ ms and a packet dropping ratio below $1$\%. Likewise, Table \ref{tab:delay_drop_bounds_2} shows the statistics obtained for $P_s=2097144$ bits and $\lambda=1/30\cdot T_t$ over the same \texttt{ULTRA} dense scenario, computed from the scatter plot in Fig. \ref{fig:fig4scatter}(b). Thresholds have been adjusted for each table discussed in this section to ensure that meaningful statistics are produced for comparison.


These tables reveal interesting insights: when dealing with small-sized packets (low $P_s$) arriving at the queues of the vTx at a high rate (high $\lambda$) the \texttt{WAF} dominates under loose constraints on the packet dropping ratio (i.e. $10$\%), whereas it is the \texttt{PSO} approach which is the outperforming method as the restriction on the number of dropped packets becomes more stringent. This changing behavior can be explained by the side benefit derived from the beamwidth optimization performed in \texttt{PSO}: narrower beamwidths would penalize the overall delay (but this penalty is restricted to the first transmission slot of every scheduling period) whereas allocating wider beamwidths make the mmWave channel more resilient against misalignments between already paired vehicles. This ultimately yields lower dropping statistics, as reflected in the table. 

A similar observation can be drawn from the statistics obtained for $P_s=2097144$ bits and $1/\lambda=1/30\cdot T_t$ packets/s. In general \texttt{PSO} outperforms the rest of the baselines in the benchmark. Nonetheless, an interesting transition is noted for average delay bounds below $0.1$: \texttt{WAF} becomes the dominating scheme and the performance of \texttt{PSO} degrades significantly. The reason for this effect is that a high value of $1/\lambda$ yields long times between transmission events, hence a lower probability that packets are dropped for all schemes in the benchmark. However, once a packet arrives at an empty queue, it takes more time to flush it through the mmWave channel due to their bigger size. It follows that, for low delay thresholds narrow beamwidths are more effective for delivering the packet to its destination, disregarding whether they are suboptimal for the delay of the scheduling slot. Indeed the \texttt{PSO} scheme fails to meet a minimum average delay of $0.05$ ms for any of its scheduling periods, as opposed to the rest of schemes (all of them with $5^\circ$ beamwidth), for which the \texttt{WAF} scheme meets this bound with a packet dropping ratio below $0.1$\% in more than $49$\% of its scheduling intervals.

\section{Conclusions and Future Research Directions} \label{sec:conclusions}

This paper has presented a novel distributed association and beam alignment framework for mmWave V2V networks based on matching theory and swarm intelligence. Specifically we have formulated tailored utility functions for the matching game that capture 1) the relative dynamics between vTxs and vRxs in the scenario; 2) the channel and queuing dynamics learned from the past and 3) the particularities of mmWave communications, such as directionality, blockage and alignment delay. This set of utilities is fed to a deferred acceptance algorithm, which allows for pairing transmitting and receiving vehicles in a distributed manner. The matching-based association is followed by an optimization procedure that allocates transmit and receive beamwidths for each established V2V link. Simulation results confirm the expected good performance of our framework over a comprehensive number of configurations for a highway multi-lane scenario with varying vehicle densities. 

Future research will be directed towards assessing the performance of this hybrid approach in multi-vUE configurations and in non-linear road networks subject to more likely misalignments between vehicles. In particular we will delve into the interplay among the scheduling period, the packet arrival statistics and the density of vehicles in real scenarios, for which we expect that the beamwidth optimization presented in this research work will render notable performance gains.

\bibliographystyle{IEEEtran}
\bibliography{IEEEabrv,jsac_mmW}
\vspace{-1cm}
\begin{IEEEbiography}[{\includegraphics[width=1in,clip,keepaspectratio]{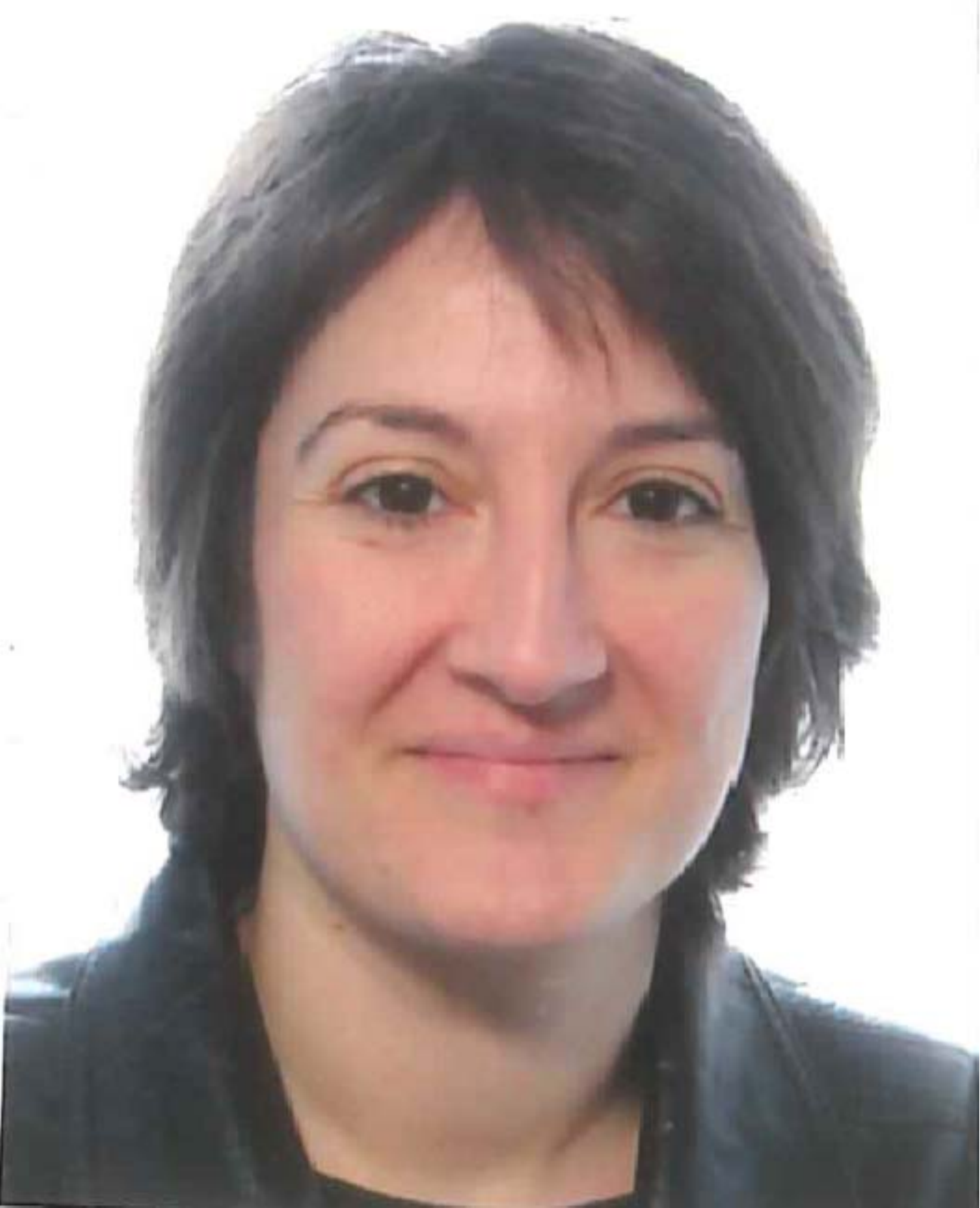}}]{Cristina Perfecto} received her B.Sc. and M.Sc. in Telecommunication Engineering by the University of the Basque Country (UPV/EHU) in 2000. She is currently an associate professor at the Department of Communications Engineering of this same University, where she teaches computer network architectures, protocols and the deployment and convergence of communications networks. Her current research interests lie on machine machine learning and data analytics including different fields such as metaheuristics and bio-inspired computation, both from a theoretical and applied point of view. She is currently working towards her Ph.D. focused on the application of multidisciplinary computational intelligence techniques in radio resource management for 5G.
\end{IEEEbiography}
\vspace{-1cm}
\begin{IEEEbiography}[{\includegraphics[width=1in,clip,keepaspectratio]{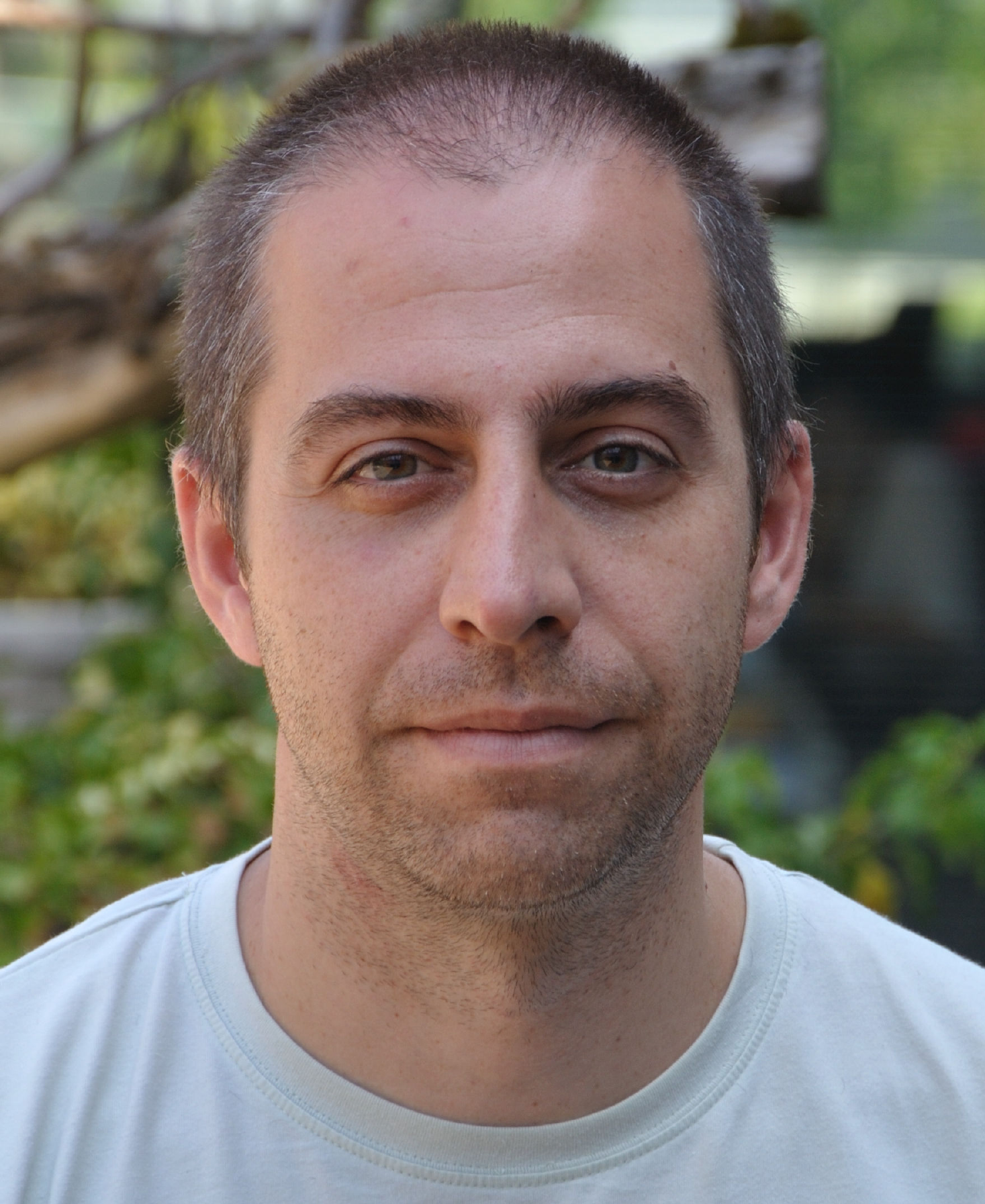}}]{Javier Del Ser} (Senior Member, IEEE) received his first PhD in Telecommunication Engineering (Cum Laude) from the University of Navarra, Spain, in 2006, and a second PhD in Computational Intelligence (summa Cum Laude) from the University of Alcala, Spain, in 2013. He is currently a principal researcher in data analytics and optimization at TECNALIA (Spain), an associate researcher at the Basque Center for Applied Mathematics and an adjunct professor at the University of the Basque Country (UPV/EHU). His research activity gravitates on the use of descriptive, prescriptive and predictive algorithms for data mining and optimization in a diverse range of application fields such as Energy, Transport, Telecommunications, Health and Security, among many others. In these fields he has published more than 160 publications, co-supervised $6$ Ph.D. theses, edited $3$ books, coauthored $6$ patents and led more than 35 research projects. Dr. Del Ser has been awarded the \emph{Talent of Bizkaia} prize for his curriculum. 
\end{IEEEbiography}
\vspace{-1cm}
\begin{IEEEbiography}[{\includegraphics[width=1in,clip,keepaspectratio]{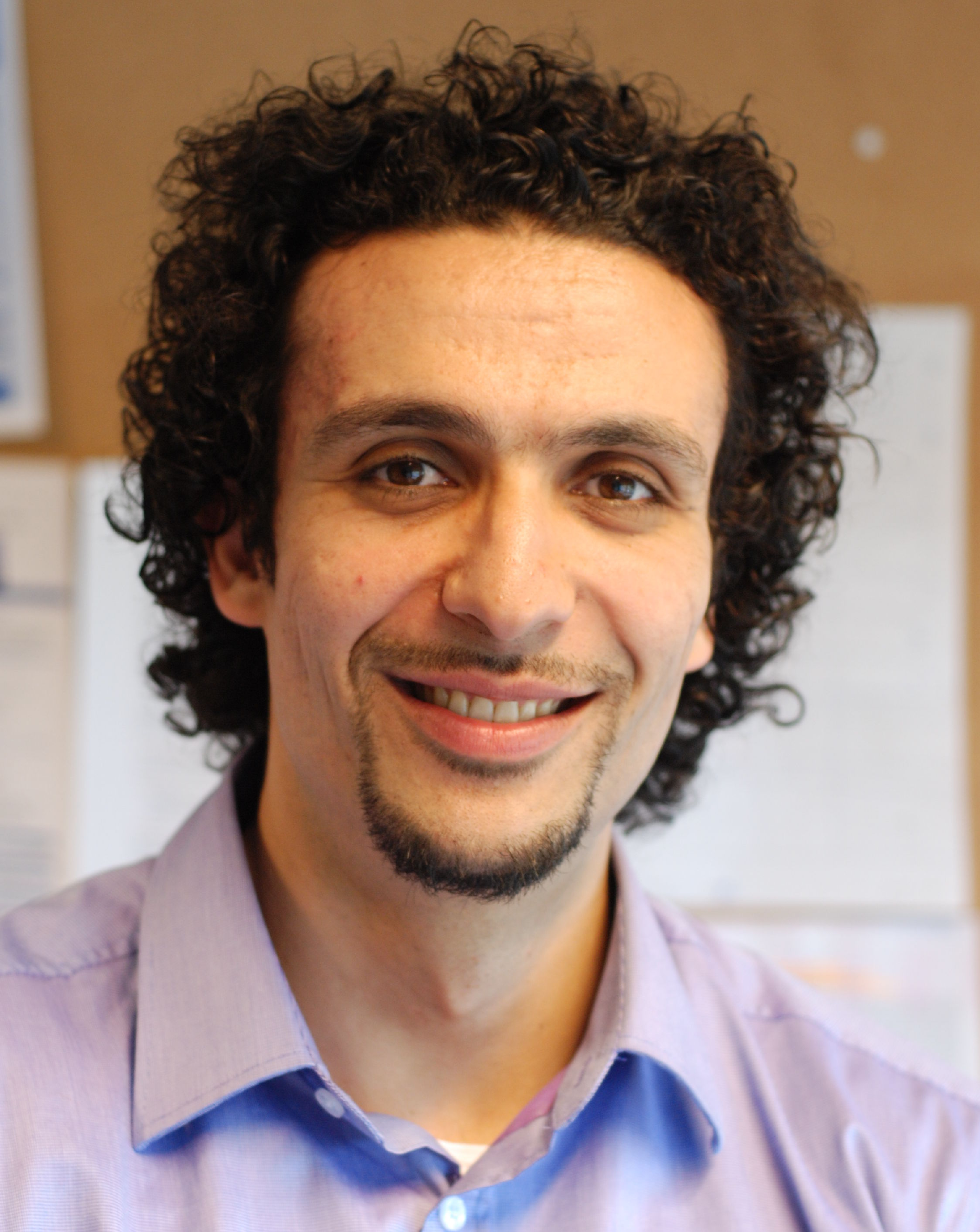}}]{Mehdi Bennis} (Senior Member, IEEE) received his M.Sc. degree in Electrical Engineering jointly from the EPFL, Switzerland and the Eurecom Institute, France in 2002. From 2002 to 2004, he worked as a 
research engineer at IMRA-EUROPE investigating adaptive equalization algorithms for mobile digital TV. In 2004, he joined the Centre for Wireless Communications (CWC) at the University of Oulu, Finland as a research scientist. In 2008, he was a visiting researcher at the Alcatel-Lucent chair on flexible radio, SUPELEC. He obtained his Ph.D. in December 2009 on spectrum sharing for future mobile cellular systems. Currently Dr. Bennis is an Adjunct 
Professor at the University of Oulu and Academy of Finland research fellow. His main research interests are in radio resource management, heterogeneous networks, game theory and machine learning in 5G networks and beyond. He has co-authored one book and published more than 100 research papers in international conferences, journals and book chapters. He was the recipient of the prestigious 2015 Fred W. Ellersick Prize from the IEEE Communications Society, the 2016 Best Tutorial Prize from the IEEE Communications Society and the 2017 EURASIP Best paper Award for the Journal of Wireless Communications and Networks. Dr. Bennis serves as an editor for the IEEE Transactions on Wireless Communication.
\end{IEEEbiography}
\vfill
\end{document}